\documentclass[useAMS,usenatbib]{mn2e}

\usepackage{graphicx}

\newcommand       \be           {\begin{equation}}
\newcommand       \ee           {\end{equation}}
\newcommand       \bea          {\begin{eqnarray}}
\newcommand       \eea          {\end{eqnarray}}
\newcommand       \apj          {ApJ}
\newcommand       \apjl         {ApJL}
\newcommand       \aap          {A\&A}
\newcommand       \nat          {Nature}
\newcommand       \mnras        {MNRAS}
\def\simlt{\mathrel{\hbox{\rlap{\hbox{\lower4pt\hbox{$\sim$}}}\hbox{$<$}}}}
\def\simgt{\mathrel{\hbox{\rlap{\hbox{\lower4pt\hbox{$\sim$}}}\hbox{$>$}}}}
\def\lesssim{\mathrel{\hbox{\rlap{\hbox{\lower4pt\hbox{$\sim$}}}\hbox{$<$}}}}

\def\gtrsim{\mathrel{\hbox{\rlap{\hbox{\lower4pt\hbox{$\sim$}}}\hbox{$>$}}}}

\title[Time-dependent Models of Accretion Disks Formed from Compact Object Mergers]{Time-dependent models of accretion disks formed from compact object mergers} \author[B.~D. Metzger, A.~L. Piro, E. Quataert]{B.~D. Metzger\thanks{E-mail:
bmetzger@astro.berkeley.edu}, A.~L. Piro, and E. Quataert \\
Astronomy Department and Theoretical Astrophysics Center,
University of California, Berkeley, 601 Campbell Hall, Berkeley CA,
94720\\}
\begin{document}
\date{Accepted . Received ; in original form }
\pagerange{\pageref{firstpage}--\pageref{lastpage}} \pubyear{????}
\maketitle
\label{firstpage}

\begin{abstract}
We present time-dependent models of the remnant accretion disks
created during compact object mergers, focusing on the energy
available from accretion at late times and the composition of the disk
and its outflows.  We calculate the dynamics near the outer edge of
the disk, which contains the majority of the disk's mass and
determines the accretion rate onto the central black hole. This
treatment allows us to follow the evolution over much longer
timescales ($100\ {\rm s}$ or longer) than current hydrodynamic
simulations.  At late times the disk becomes advective and its
properties asymptote to self-similar solutions with an accretion rate $\dot{M}_{d} \propto
t^{-4/3}$ (neglecting outflows).  This late-time accretion can in
principle provide sufficient energy to power the late-time activity
observed by {\it Swift} from some short-duration gamma-ray bursts.  However, because outflows during the advective phase unbind the majority of the
remaining mass, it is difficult for the remnant disk alone to produce
significant accretion power well beyond the onset of the advective
phase.  Unless the viscosity is quite low ($\alpha\lesssim10^{-3}$),
this occurs before the start of observed flaring at $\sim 30$ s;
continued mass inflow at late times thus appears required to explain
the late-time activity from short-duration gamma-ray bursts.  We show that the composition of the disk freezes-out when the disk is
relatively neutron rich (electron fraction $Y_e\simeq 0.3$). Roughly
$10^{-2} M_\odot$ of this neutron-rich material is ejected by winds at
late times.  During earlier, neutrino-cooled phases of accretion, neutrino
irradiation of the disk produces a wind with $Y_e\simeq0.5$, which synthesizes at
most $\sim10^{-3}M_\odot$ of $^{56}$Ni.  We highlight what conditions
are favorable for $^{56}$Ni production and predict, in the best cases,
optical and infrared transients peaking $\sim 0.5-2$ days after the
burst, with fluxes a factor of $\sim 10$ below the current
observational limits.
\end{abstract}


\section{Introduction}
\label{sec:int}
\voffset=-2cm

\vspace{0.2 cm}
\label{sec:introduction}
The most popular model for the creation of short duration gamma-ray
bursts (GRBs) is either binary neutron star (NS/NS) or black
hole-neutron star (BH/NS) coalescence \citep{pac86,pac91,eic89,nar91}.
Support for the merger hypothesis comes from their durations of $\lesssim2\ {\rm s}$, observations of well-localized short GRBs in galaxies
without strong star formation
\citep{ber05,ger05,2005ApJ...630L.117H}, and the lack of a detectable
coincident supernovae \citep{2005ApJ...630L.117H,blo06,sod06,fer07}, as
is found in the case of long ($\gtrsim2\ {\rm s}$) GRBs
\citep{gal98,hjo03,sta03}.

Previous theoretical studies of the merger process have focused on one
of two stages. The first is the {\it dynamical} portion in which the less massive companion is tidally disrupted by the more
massive BH \citep{lk95,lk98,lk99,kl98,jan99,ros04} or NS
\citep{ruf96,rj99,oj06}. The details of whether a dynamical
instability \citep{rs94,lai94} or Roche lobe overflow occurs depends
on the mass ratio and the nuclear equation of state
\citep{bc92,ue99}.  

   Nevertheless, generally $\sim~0.01-0.1M_\odot$ of material remains in a remnant disk following the dynamical stage.  The accretion of this material onto the central object gives rise to the second, {\it
disk} portion of the merger.  The energetics and timescale of the accretion phase are reasonably consistent with observations of short GRBs, as
was shown by models of steady-state, azimuthally symmetric, vertically
averaged disks \citep{pop99,nar01,km02,dim02,cb07}.  More recently,
these disks have been modeled with time-dependent calculations in 1D
\citep{jan04}, 2D \citep{lee04,lee05b}, and 3D \citep{set04,set06}.  The typical time interval that present multi-dimensional calculations can simulate is on the order of the burst duration or less ($\sim1-2\ {\rm s}$ for 2D and $\sim50\ {\rm ms}$ for 3D).  

Recent observations of short GRBs by {\it Swift}, however, indicate continued activity from the central engine on much longer timescales.  X-ray flares with durations of $\sim100\ {\rm s}$ after a delay of $\sim30\ {\rm s}$
have been seen from several bursts
\citep{bar05,vil05,cam06,lap06}. Stacked lightcurves of many bursts
indicate continued activity on a similar timescale
\citep{laz01,mon05}. In one extreme case, GRB 050724 displayed an
X-ray flare $12$ hours post-burst. This flaring activity has been
attributed to a number of different sources, including fragmentation
of a rapidly rotating core \citep{kin05}, magnetic regulation of the
accretion flow \citep{pz06}, fragmentation of the accretion disk
(Perna et al. 2005; although this explanation may have difficulty
reproducing the observed timescales, Piro \& Pfahl 2007), differential
rotation in a post-merger millisecond pulsar \citep{dai06}, and an
infalling tidal tail of material stripped from the disrupted NS (Lee $\&$ Ramirez-Ruiz 2007; Rosswog 2007).

In order to determine whether the late-time activity from short GRBs is consistent with a compact merger origin, the disk evolution should be followed for timescales much longer than the initial viscous time. With this aim, we perform time-dependent calculations modeling
the disk as an annulus that contains the majority of the mass.  This
simplification allows us to study the disk evolution for arbitrarily
long timescales, and to readily determine important properties such as
the disk's composition and when it becomes advective. We are
also able to survey much of the parameter space of initial disk mass
and angular momentum. In \S \ref{sec:initial} we discuss the initial
conditions for disks formed from compact object mergers. This is
followed by \S \ref{sec:physics}, in which we summarize the main
assumptions of our ring model.  In \S \ref{sec:results} we present the
results of our calculations and summarize the main properties of the
models.  We then calculate outflows from our disk solutions in \S
\ref{sec:winds}. We investigate the composition of the outflows and
argue that they generally consist of neutron-rich isotopes, but can
produce $^{56}$Ni in some circumstances.  The presence or lack of an
optical transient from short GRBs therefore provides an important
constraint on progenitor models.  We conclude in \S \ref{sec:theend}
with a discussion of our results. In Appendix A we summarize the
Green's function solution to the viscous spreading of a ring, which is
important for connecting our ring model to the true extended disk
geometry. In Appendix B we present analytic self-similar solutions
that reproduce many of the features of our numerical solutions.

\section{Initial Conditions}
\label{sec:initial}

   The dynamical phase of NS/NS or BH/NS mergers has
been studied extensively using a number of different numerical
techniques and methods for including general
relativity (GR).  Here we summarize some of the most relevant features for our study
\citep[for a more detailed review, see][]{lr07}.  

   When the lighter companion NS is first tidally disrupted, a debris disk is
formed within only a few dynamical timescales.  The initial disk mass, $M_{d,0}$, is generally larger for more
asymmetric mass ratios (i.e., small $q$, where $q$ is the ratio of the
lighter to the heavier binary component). For example, \citet{st06}
find that for a NS/NS merger with $q=0.7$ that $M_{d,0}=0.03\
M_\odot$, but for $q=0.9$ the disk is much less massive with
$M_{d,0}=10^{-3}\ M_\odot$.  Another trend is that including strong
gravity gives less massive remnant disks.  The BH spin is also important, with larger spin favoring disk formation \citep{ras05}
and the production of a tidal tail. These have masses of
$\simeq0.01-0.05\ M_\odot$ and may provide prolonged mass inflow
\citep{lr07}, but for simplicity this will be ignored here.  Taken 
together, these simulations generally find $M_{d,0}\simeq0.01-0.3M_\odot$, with the disk containing a substantial fraction of the angular momentum of the disrupted companion.

In the standard picture of NS-NS mergers, the resulting hypermassive NS collapses to a BH shortly following the merger.  However, simulations show that when (and if) collapse actually occurs depends on the mass of the central NS and its ability to transport angular momentum to the surrounding disk (Shibata et al.~2005; Shibata $\&$ Taniguchi 2006; Shibata et al.~2006).  In fact, if the NS remains supported by differential rotaton for several seconds (Baumgarte et al.~2000; Morrison et al.~2004; Duez et al.~2004, 2006) or loses sufficient mass via a centrifugally-driven outflow (e.g., Thompson et al.~2004; Dessart et al.~2008a), the NS itself may power the GRB (e.g., Price $\&$ Rosswog 2006).  In this paper we assume that the central object promptly collapses to a BH; our model, however, would be reasonably applicable for the case of a central NS as well, the primary difference being that the significant neutrino flux from the newly-formed NS and from the boundary layer between the disk and the NS could modify the composition and thermal properties of the disk.

We present some characteristic numbers to motivate our choice of
initial conditions.  Consider a binary with masses $M$ and $m$
($M>m$), where the latter is the NS (with radius $R$) that is tidally
disrupted. The disruption radius, $a_t$, is estimated to be (Kopal
1959, adding Fishbone's 1973 10\% strong gravity correction) \be
a_t\simeq2.4R \left(\frac{M+m}{m}\right)^{1/3}.  \ee The characteristic
orbital period at this radius is \be P_t \simeq
23.4\left(\frac{R^3}{Gm}\right)^{1/2} \simeq
2\times10^{-3}m_{1.4}^{-1/2}R_6^{3/2}\ {\rm s}, \ee where
$m_{1.4}=m/1.4\ M_\odot$ and $R_6=R/10^6\ {\rm cm}$, with an orbital
angular momentum of \bea J_t &=&  \left( G(M+m)a_t\right)^{1/2}m \nonumber \\
&\simeq& 6\times10^{49}(1/q+1)^{2/3}m_{1.4}^{3/2}R_6^{1/2}\ {\rm ergs\
s}, \eea where $q=m/M$. The disrupted NS also contains spin angular
momentum. This is negligible since the NS is not strongly affected by
tidal coupling \citep{bc92}.  Even a rapidly rotating NS ($\simeq5\
{\rm ms}$) has an associated angular momentum of merely $\sim10^{48}\
{\rm ergs\ s}$.

   Once disrupted, a considerable fraction of the NS is either lost
from the system or immediately swallowed by the BH. The remaining material forms a thick torus surrounding the central BH. Its associated viscous
timescale can be estimated by assuming that the majority of the torus'
mass lies at a single radius, $r_{d,0}$. Taking the angular momentum
of the disk to be $J_d\simeq(GMr_{d,0})^{1/2}M_{d,0}$, we estimate \be
r_{d,0}\simeq3\times10^7M_3^{-1}M_{0.1}^{-2}\left(\frac{J_{49}}{2}\right)^2\
{\rm cm},
	\label{eq:rd0}
\ee
where $M_3= M/3\ M_\odot$, $M_{0.1}=M_{d,0}/0.1\ M_\odot$, and
$J_{49}=J_d/10^{49}\ {\rm ergs\ s}$.
For a disk with half-thickness $H$, the viscous timescale is
\bea
	t_{\rm visc,0}& = &\alpha^{-1}\left(\frac{r_d}{H}\right)^2 \left(\frac{r_d^3}{GM}\right)^{1/2}
	\nonumber
	\\ &\simeq &6\times10^{-2}\alpha_{0.1}^{-1}
	M_3^{-1/2}r_7^{3/2}\left(\frac{H}{0.5r_d}\right)^{-2}\ {\rm s},
	\label{eq:tvisc0}
\eea where $\alpha = 0.1\alpha_{0.1}$ is the standard
dimensionless viscosity \citep{ss73}, $r_7=r_{d,0}/10^7\ {\rm cm}$,
and we have scaled to an initial ratio of $H/r_d=0.5$, consistent with
our numerical solutions.  The initial viscous time $t_{\rm visc,0}$ is
roughly the time at which the central BH begins accreting in earnest.
The strong dependence of $t_{\rm visc,0}$ on disk mass and radius demonstrates that the initial evolution of the disk is sensitive to the
outcome of the dynamical phase of the merger. But as we will show, the
late time evolution is much less sensitive to initial conditions and
is well described by self-similar solutions.

\section{Physics of the Expanding Ring Model}
\label{sec:physics}

   Given these initial conditions, one would like to know how the disk
then evolves. Modeling the entire disk requires resolving timescales
over $\sim4-6$ orders of magnitude.  This makes it expensive to carry
out simulations for long periods of time. We consider instead a
simplified model that captures most of the features of interest. At
any given time, $t$, the disk can be broken into three regions
depending on the local viscous time, $t_{\rm visc}$, which increases
with radius, roughly as $t_{\rm visc}\sim r^{3/2}$.  At small radii,
$t_{\rm visc}<t$, and the disk comes into steady-state.  This is the
region most often modeled in previous studies
\citep{pop99,nar01,km02,dim02,cb07}. The radii where $t_{\rm visc}\sim
t$ contain the majority of the disk's mass and angular
momentum. Therefore, this region determines the viscous evolution of
the rest of the disk, including the mass accretion rate that is fed to
the interior steady-state region. Motivated by this fact, we focus on
this radius and model the disk as a ring. Exterior to this point is a
third region where $t_{\rm visc}>t$, but this contains a small amount
of mass and is negligible for the viscous evolution.

\subsection{Dynamical Equations}
\label{sec:dynamical}

Our ring model treats the disk as a single annulus that is evolved
forward in time. In this picture, the properties of the ring, such as
its surface density $\Sigma$ and temperature $T$, are representative
of the location where $\Sigma r^2$ peaks.  The main drawback of this
method is that the material in the disk is in fact distributed
spatially in radius.  Thus, although the mass of the disk in the
vicinity of $r_d$ is $\simeq \pi \Sigma r_d^2$, the total mass of the
disk (integrated over all radii) is $M_d=A \pi \Sigma r_d^2$, where A
is a factor of order unity that accounts for the distinction between
the total mass of the disk and the mass of the material near $r_d$.
Similarly, we write the total angular momentum of the disk as $J_d =
B(GMr_d)^{1/2}\pi r_d^2\Sigma$.  At early times the constants $A$ and
$B$ depend on the initial conditions of how matter is spatially
distributed; however, at times much greater than the initial viscous
time (given by eq. [\ref{eq:tvisc0}]), material initially concentrated
at a given radius becomes spread out in a manner determined by the
viscosity. As described in detail in Appendix A, we choose the
constants $A$ and $B$ by setting the solution of our simplified ring
model at late times equal to the Green's function solution for a
spreading ring with a viscosity $\nu \propto r^{1/2}$ (as is
appropriate for the radiatively inefficient disk at late-times).  This
fixes $A = 3.62$ and $B = 3.24$.\footnote{In fact, when the total
angular momentum is conserved, the viscous evolution is independent of
$A/B$ as long as $A/B$ is nearly constant with time.}  Conveniently
$A/B\simeq1$, so that it is a good approximation to take
$J_d\simeq(GMr_d)^{1/2}M_d$.

   The time evolution of the disk is determined by the conservation equations.
Conservation of mass is
\be
	\frac{d}{dt}\left(A\pi\Sigma r_d^2\right)=-\dot{M}_d,
	\label{eq:mass}
\ee where $\dot{M}_d$ is in general the total mass loss rate, which could include both accretion and a wind (for now we ignore the effects of a
wind). Conservation of angular momentum is \be \frac{d}{dt}\left[
B(GMr_d)^{1/2}\pi\Sigma r_d^2\right]=-\dot{J},
	\label{eq:ang_mom}
\ee
where $\dot{J}$ is the angular momentum loss rate.
Equations (\ref{eq:mass}) and (\ref{eq:ang_mom}) provide two coupled
equations that can be solved for the dependent variables $r_d$ and $\Sigma$.

   The accretion rate must depend on the characteristic mass and
viscous timescale of the ring, so we use
\be
	\dot{M}_{d} = fM_d/t_{\rm visc},
	\label{eq:mdotacc}
\ee where $t_{\rm visc}=r_d^2/\nu$ and $\nu$ is the viscosity.  The
factor $f$ is set like $A$ and $B$ to match the exact solution of a
spreading ring with $\nu \propto r^{1/2}$ (Appendix A), which gives
$f=1.6$.\footnote{Although we set $t_{\rm
visc}=r^2/\nu$, any prefactors that could go into this prescription
would just be absorbed into a re-definition of $f$.}  Requiring a
no-torque boundary condition at a radius $r_*$, we take \be f =
1.6/[1-(r_*/r_d)^{1/2}].  \ee In contrast, a steady-state disk obeys
$\dot{M}_{d}=3\pi\nu\Sigma$ (ignoring the no-torque condition), which
instead gives $f=3/A\simeq0.83$.  

For the viscosity, we use an $\alpha$-prescription, \be \nu=\alpha
c_sH, \ee where $c_s=(P/\rho)^{1/2}$ is the isothermal sound speed.
The equation of state includes contributions from radiation pressure,
gas pressure, relativistic degeneracy pressure, and neutrino pressure
as in \citet{dim02}.

\subsection{Energetics}

For the energy equation, we take \be q_{\rm visc} = q_\nu^- +q_{\rm
adv}, \label{eq:energy} \ee where $q_{\rm visc}$ is the viscous
heating, $q_\nu^-$ is the neutrino cooling \citep[using the
prescriptions given by][which includes neutrino optical-depth
effects]{dim02}, $q_{\rm adv}$ is the advective heat flux, and all $q$
values correspond to half the disk thickness.

For a disk rotating at the Keplerian frequency $\Omega =
(GM/r_d^3)^{1/2}$, \be q_{\rm visc}=\frac{9}{8}\nu\Omega^2\Sigma =
\frac{9}{8fA}\frac{GM\dot{M}_{d}}{\pi
r_d^3}\left[1-\left(\frac{r_*}{r_d}\right)^{1/2} \right],
	\label{eq:qvisc}
\ee where the prefactor $9/(8fA)\simeq0.2$ is different from the
steady-state value of $3/8$.  The advective term, $q_{\rm adv}$, is
set as in \citet{dim02}, with the only difference being that the
radial velocity is the expansion rate of the ring's radius \be V_r =
\frac{dr_d}{dt} = \frac{2\dot{M}}{A\pi r_d\Sigma}, \ee where we have
taken $\dot{J}=0$. 

Fusion to $\alpha$-particles produces heating in addition to
$q_{\rm visc}$, with \be q_{\rm
nucl}=6.8\times10^{28}\rho_{10}\frac{dX_{\rm \alpha}}{dt}H, \ee where
all quantities are expressed in cgs units, $\rho_{10}=\rho/10^{10}\
{\rm g\ cm^{-3}}$ and $X_{\rm \alpha}$ is the mass fraction of
$\alpha$-particles. Note that in our case $q_{\rm nucl}>0$ since
$\alpha$-particles are synthesized as the disk expands (in contrast to
studies that follow cooling from photodisintegration as material moves
inward).  In our calculations we do not include $q_{\rm nucl}$ in solving
equation (\ref{eq:energy}) because we were not able to find reasonable
solutions when doing so (for reasons explained in \S \ref{sec:gross}).

\subsection{Composition}

   An advantage of the ring model is that other properties of the
disk, such as its composition, can be cast into differential equations
and integrated along with equations (\ref{eq:mass}) and
(\ref{eq:ang_mom}).  Since the neutron content of the disk is
particularly important for determining the properties of the disk's
outflows, we evolve the electron fraction $Y_e$ using \be
\frac{dY_e}{dt} = -Y_e r_{e^-p}+(1-Y_e) r_{e^+n},
	\label{eq:ye}
\ee
where $Y_e=X_p/(X_n+X_p)$, $X_p$ and $X_n$ are the proton and
neutron mass fraction, respectively, and $r_{e^-p}$ and $r_{e^+n}$ are
the electron and positron capture rates, respectively (Beloborodov 2003a).  We have neglected the effect of neutrino absorptions on the evolution of $Y_{e}$ in equation (\ref{eq:ye}).  Although absorptions are important at early times when the disk is optically thick, we are primarily concerned with the late-time value of $Y_{e}$, which does not depend sensitively on the neutrino irradiation (see $\S\ref{sec:composition}$). 

   As the disk evolves, the protons and neutrons eventually burn to
form $\alpha$-particles. At these times the disk is sufficiently cold that
the positron and electron capture rates are negligible
(i.e, $1/r_{e^-p}\gg t_{\rm visc}$) and $Y_e$ has frozen-out. This fixes the
difference between the free neutron and proton mass fractions
\be
	X_n-X_p=1-2Y_e.
	\label{eq:frozenratio}
\ee
Since the rates for reactions that synthesize and destroy $\alpha$-particles
are all fast in comparison
to the viscous time, we determine the composition using
nuclear statistical equilibrium (NSE) between protons, neutron, and $\alpha$-particles.
This is expressed by the Saha relation \citep{st83}
\be
	X_p^2X_n^2
	= 1.57\times10^4\ X_\alpha \rho_{10}^{-3}
	T_{10}^{9/2}
	\exp\left(-\frac{32.81}{T_{10}} \right).
	\label{eq:saha}
\ee
NSE is a good assumption because the disk temperature is generally $\gtrsim 0.5$ MeV (see Fig.~\ref{fig:temperature}), except at very late
times or for very low disk masses (e.g., the $M_{d,0} = 0.03M_{\sun}$ case,   
for which we do not calculate the nuclear composition anyways).  By combining equations (\ref{eq:frozenratio}) and (\ref{eq:saha}) with mass conservation,
$X_p+X_n+X_\alpha=1$, we solve for all of the mass fractions at a given
$\rho$, $T$, and $Y_e$.

\section{Time-Evolving Solutions}
\label{sec:results}

   We next present the results of integrating equations
(\ref{eq:mass}), (\ref{eq:ang_mom}), and (\ref{eq:ye}) forward in
time.  For simplicity, we typically assume that $\dot{J} = 0$.
A convenient property of our formalism is the ease with which these
complications can be included (for example, we consider the effects of winds at the end
of \S 4.1). The disk
properties are determined by the initial conditions $M_{d,0}$, $J_d$,
and $Y_{e,0}$, and by the viscosity $\alpha$.  For the majority of our
study we set the initial $Y_{e,0}=0.1$, which is characteristic of the
inner neutron star crust \citep{hz90a,hz90b,pr95}. An additional
important parameter is $r_*$, which is set by the spin of the central
BH.  In most of our calculations we take $r_*\simeq2.3r_g\simeq1.02\times10^6\ {\rm cm}$, corresponding to the innermost stable circular orbit of a $3\ M_\odot$ BH with spin $a\simeq0.9$; when calculating the properties of disk outflows in $\S\ref{sec:winds}$, however, we also consider the case of a nonrotating ($a=0$) BH.  We consider the general evolution of the disk in \S
\ref{sec:gross}, and then focus on the composition in \S
\ref{sec:composition}.

\begin{figure}
\resizebox{\hsize}{!}{\includegraphics[ ]{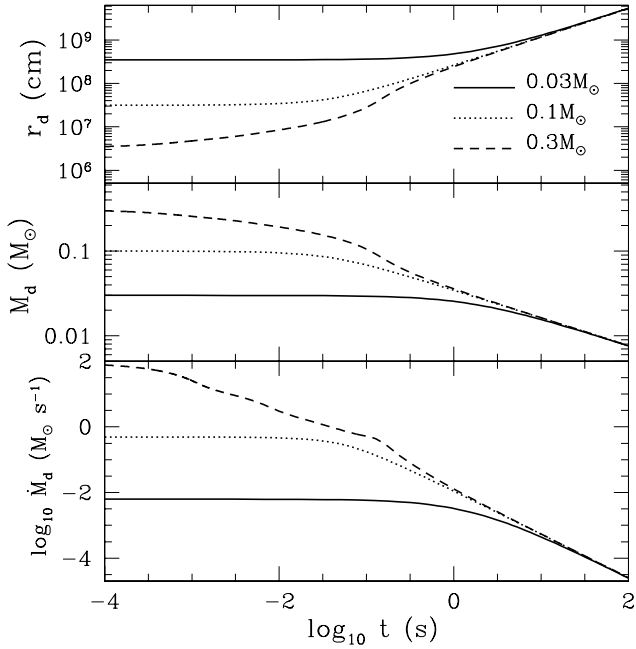}}
\caption{Example disk models showing the evolution of the disk radius,
$r_d$, disk mass, $M_d$, and accretion rate, $\dot{M}_{d}$, as a
function of time.  We compare $M_{d,0}=0.03$ ({\it solid lines}),
$0.1$ ({\it dotted lines}) and $0.3\ M_\odot$ ({\it dashed lines})
solutions; all use $J_{49}=2$ and $\alpha=0.1$. The inner radius is
$r_*\simeq2.3r_g\simeq1.02\times10^6\ {\rm cm}$ (corresponding to a
$3\ M_\odot$ BH with a spin of $a\simeq0.9$).
}
\label{fig:mass1}
\end{figure}

\begin{figure}
\resizebox{\hsize}{!}{\includegraphics[ ]{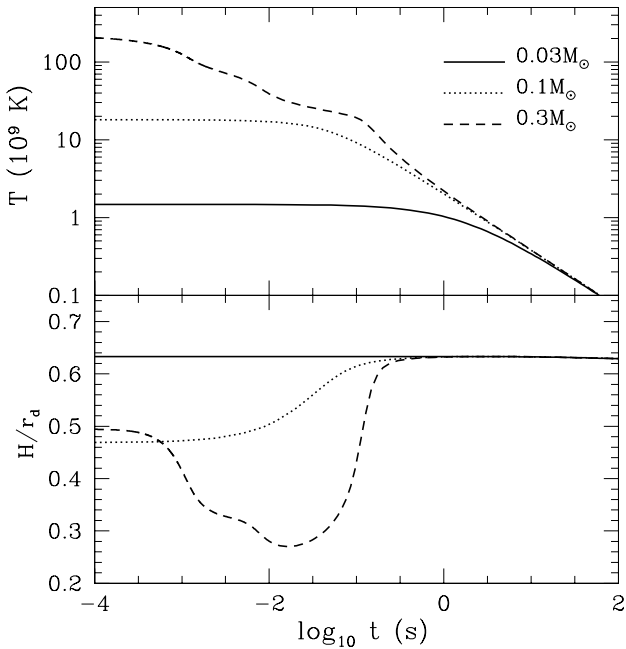}}
\caption{Comparison of the midplane temperatures and scaleheight for the three
models from Fig. 1. In the lowest mass model, the ring
is always advectively-dominated, thus $H/r_d$ is constant.}
\label{fig:temperature}
\end{figure}

\subsection{Disk Evolution and Energetics}
\label{sec:gross}

    At any given time, a ring model is in one of three phases: (1)
early-time, optically thick to neutrinos and advectively dominated,
(2) mid-time, optically thin to neutrinos and geometrically thin, and
(3) late-time, radiatively-inefficient accretion flow
(RIAF).\footnote{An optically thick, geometrically thin stage occurs
between stages (1) and (2); however, this phase is brief and is not
dynamically very different from phase (2), so we do not consider it
separately in our discussion.}  This is analogous to the different
regions of steady-state, hyper-accreting accretion disks \citep[see,
e.g.,][]{cb07}, but now the transitions occur with time instead of
radius. The phases that a certain ring model samples during the course
of its viscous expansion depends on $t_{\rm visc,0}$. A more compact
disk (a shorter $t_{\rm visc,0}$) will exhibit all three phases, while
larger disks may only exhibit phases (2) and (3), or even just (3).

\begin{figure}
\resizebox{\hsize}{!}{\includegraphics[ ]{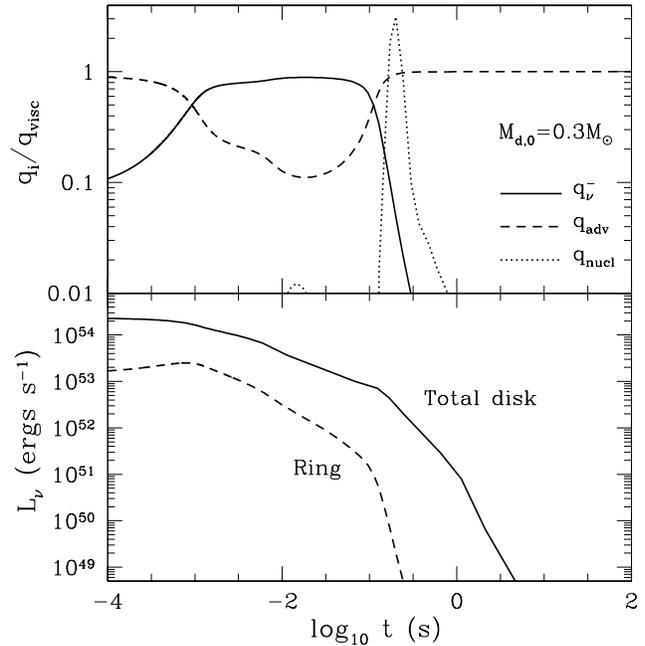}}
\caption{The cooling rates and neutrino luminosity
for the $M_{d,0}=0.3M_\odot$ model from Fig. \ref{fig:mass1}. For the
cooling rates we compare the neutrino ({\it solid line}) and
advective ({\it dashed line}) rates, normalized to the viscous
heating. The implied heating from the creation of $\alpha$-particles
is plotted as a dotted line, but is not accounted for in the disk
evolution.  The neutrino luminosities are from the entire disk ({\it
solid line}) and the ring ({\it dashed line}). The former luminosity
is estimated by integrating over a steady-state disk model at each
time given $\dot{M}_{d}(t)$.}
\label{fig:energy1}
\end{figure}

\begin{figure}
\resizebox{\hsize}{!}{\includegraphics[ ]{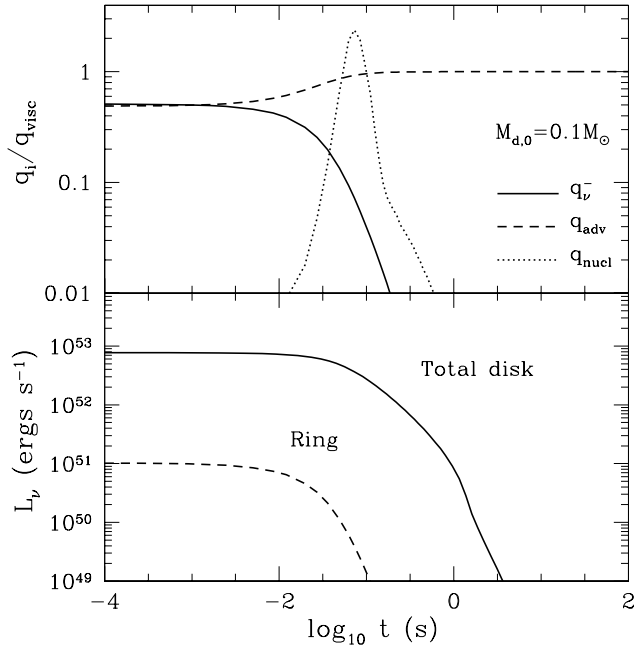}}
\caption{The same as Fig. \ref{fig:energy1}, but for
$M_{d,0}=0.1M_\odot$.}
\label{fig:energy2}
\end{figure}

\begin{figure}
\resizebox{\hsize}{!}{\includegraphics[ ]{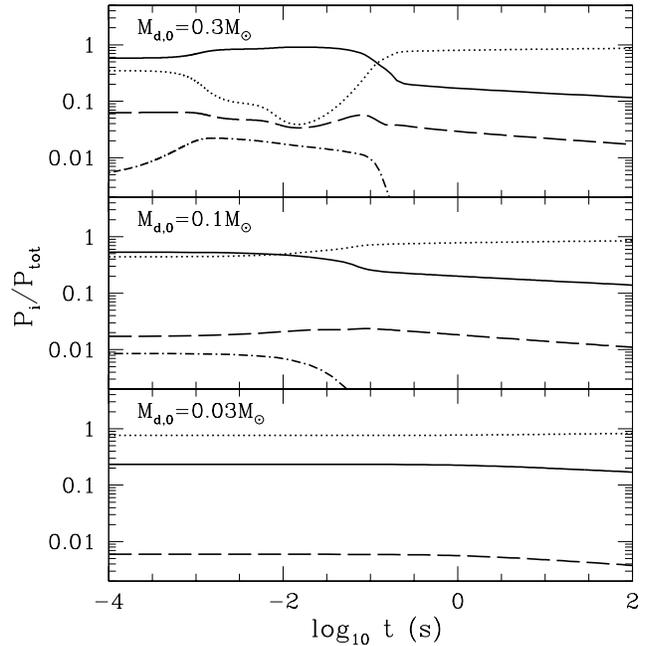}}
\caption{Pressure contributions for $M_{d,0}=0.3M_\odot$ ({\it top
panel}), $0.1M_\odot$ ({\it middle panel}) and $0.03M_\odot$ ({\it
bottom panel}). The pressures are all normalized to the total pressure
and include the ion pressure ({\it solid lines}), radiation pressure
({\it dotted lines}), degenerate electron pressure ({\it dashed
line}), and neutrino pressure ({\it dot-dashed line}).}
\label{fig:pressure}
\end{figure}

    We present a number of figures that are helpful in understanding
these three phases and how they are affected by changing $M_{d,0}$.
Figure \ref{fig:mass1} shows the radius $r_d$, mass $M_d$, and
accretion rate $\dot{M}_{d}$ as a function of time, for $M_{d,0} =
0.3, 0.1,$ and $0.03 \, M_\odot$.  Figure \ref{fig:temperature} compares the
midplane temperature and scaleheight for these same models.
Figures \ref{fig:energy1} and
\ref{fig:energy2} show key results describing the energetics of the
$M_{d,0}=0.3$ and $0.1\ M_\odot$ solutions, respectively, while Figure
\ref{fig:pressure} shows the different contributions to the total
pressure in the disk as a function of time. Note that we fix the total
angular momentum in these calculations ($J_{49}=2$) and thus a larger
$M_{d,0}$ corresponds to a smaller $r_{d,0}$ and a shorter $t_{\rm visc,0}$.

   The first transition the disks make is from an optically thick,
advective disk to a thin, neutrino-cooled disk; i.e., from phase (1)
to (2). This is only exhibited by the $M_{d,0}=0.3M_\odot$ model and is seen most clearly at early times in Figure \ref{fig:temperature}
when $H/r_d\simeq0.5$ and in Figure \ref{fig:energy1} when $q_{\rm adv}\gg q_\nu^-$.  Figure
\ref{fig:pressure} shows that this phase is ion pressure (ideal gas)
dominated.  A simple estimate determines what initial disk mass is
required for phase (1) to occur, i.e., for the initial disk to be
optically thick and advective. The disk is advective for radii inside
of which the neutrino diffusion time out of the disk exceeds the
inflow time.  Setting this radius equal to the initial radius of the
disk (eq. [\ref{eq:rd0}]), we find that there is a critical disk mass
below which the disk never experiences phase (1), 
\be
M_{d,\rm crit}\sim0.2\alpha_{0.1}^{-1/10}
M_3^{-7/10}\left(\frac{J_{49}}{2}\right)^{9/10}\left(\frac{H}{0.5r_d}\right)^{-3/5}M_\odot,
	\label{eq:mcrit}
\ee where we have dropped scalings with $f$ and $A$ since they appear
raised to the $1/10$ power. This estimate is consistent with the fact that
our $M_{d,1}=0.1M_\odot$ model is not advective at early times, as
seen in Figures \ref{fig:temperature} and \ref{fig:energy2}.  In this case only phases (2) and (3) are seen,
i.e., the disk is initially thin and neutrino cooled and later
transitions to being advective.

   Once the models reach the late-time, RIAF phase, or phase (3), they
asymptote to self-similar solutions, independent of the initial disk
mass.  In this phase, the disk has $q_{\rm adv}>q_\nu^-$ and is
radiation pressure dominated.  We derive analytic self-similar
solutions in Appendix B2 for this limit and show that $r_d \propto
t^{2/3}$, $M_d \propto t^{-1/3}$ and $\dot M_{d} \propto t^{-4/3}$.  The
RIAF solution occurs external to an ``ignition radius,'' which we estimate as the location where the pair capture cooling rate balances $\sim 1/2$ of the viscous heating for a thick disk,
\bea
r_{\rm ign}\simeq \nonumber \eea \bea 3\times10^7\alpha_{0.1}^{-2}M_3^{-3/5}\left(\frac{H/r_{d}}{0.4}\right)^{-14/5} \left(\frac{\dot{M}_{d}}{0.1M_\odot{\rm s^{-1}}}\right)^{6/5} {\rm cm},
	\label{eq:rign}
\eea
where we have scaled $H/r_{d}$ to $\approx 0.4$, a value appropriate for the transition between the thin and thick disk regimes.  We combine this with the analytic results for $r_d(t)$ and
$\dot{M}_{d}(t)$ in the RIAF limit (eqs. [\ref{eq:rd_analytic2}] and
[\ref{eq:mdot_analytic2}])\footnote{We use these solutions rather than
the thin-disk ones because the numerical results follow these more
closely (Fig. \ref{fig:analytic}).}  to estimate the time when the
disk transitions to being thick, which yields \be t_{\rm
thick}\sim0.1\alpha_{0.1}^{-23/17}M_3^{-13/17}\left(\frac{J_{\rm
49}}{2}\right)^{9/17} {\rm s}.
	\label{eq:tthick}
\ee Equation (\ref{eq:tthick}) is only applicable if the disk is thin
at early times.  For sufficiently small initial disk masses, less than
\be M_{d,\rm
thick}\sim0.1\alpha_{0.1}^{2/17}M_3^{-7/17}\left(\frac{J_{\rm
49}}{2}\right)^{14/17} M_\odot, \ee this is no longer true, and the disk
is always a RIAF at its outer radius.

Figures \ref{fig:energy1} and \ref{fig:energy2} show that at
approximately the same time as the disk transitions from being thin to
thick, protons and neutrons are fused to He.  Although the nuclear
heating rate $q_{\rm nuc}$ is shown in Figures \ref{fig:energy1} and
\ref{fig:energy2}, this heating was not included in our time-dependent
calculations so that we could obtain solutions at late times.  The
nuclear heating rate is sufficiently large, i.e, $q_{\rm nucl}\gtrsim
q_{\rm visc}$, that the disk is not able to accommodate this added
energy (it is already thick with $H \simeq r$ due to viscous heating
alone). This probably implies that the burning contributes to driving
a powerful wind \citep[as described by][]{lr07}.

\begin{figure}
\resizebox{\hsize}{!}{\includegraphics[ ]{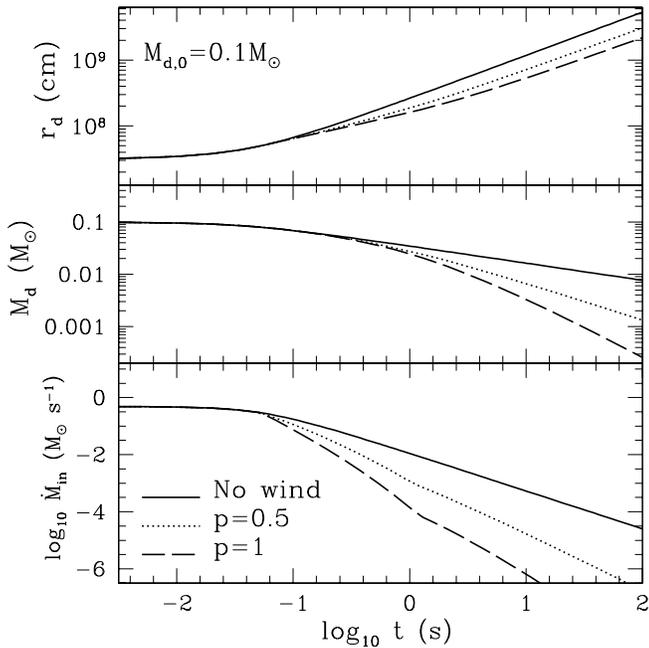}}
\caption{The radius $r_{d}$, disk mass $M_d$, and mass accretion rate reaching the central BH $\dot{M}_{\rm in}$ for different parameterizations of mass loss during the advective phase.  We initialize a disk with $M_{d,0}=0.1M_\odot$ (and all other parameters fixed as in Fig. \ref{fig:mass1}) and compare solutions with no wind ({\it solid line}), $p=0.5$ ({\it dotted line}; see eq. [B8]), and $p=1$ ({\it dashed line}).}
\label{fig:wind}
\end{figure}

However, such a wind
already begins at this time by virtue of the disk being advective (as
discussed in \S \ref{thickdiskwinds}).  In Appendix B3, we
present analytic self-similar solutions for advective disks with mass
loss and show that this significant mass loss causes $M_d$
and $\dot M_{d}$ to decline much more rapidly with time than is shown in
Figure \ref{fig:mass1}.  This is shown explicitly in Figure
\ref{fig:wind}, where we present disk models calculated using the mass
and angular momentum loss prescriptions described in Appendix B3;
such losses are assumed to occur only when the disk is thick, between
$\sim \max(r_*, r_{\rm ign})$ and $\sim r_d$.  Figure \ref{fig:wind}
compares time-dependent solutions with no wind ({\it solid line}), a
wind with $p=0.5$ ({\it dotted line}; see eq. [B8]), and a wind with
$p=1$ ({\it dashed line}).\footnote{See Appendix $\ref{sec:riaf}$ for the definition of p.}  The loss of angular momentum does not appreciably slow the radial expansion of the disk, but it does substantially accelerate the decline in the disk mass and accretion rate (see also eqs. [\ref 
{MdCne1}] and [\ref{mdotinCne1}]).  If the models with winds are  
accurate, significant accretion is only likely to last for a few  
viscous times once the disk enters the late-time advective phase.   
Continued central engine activity at much later times could result  
from late-time infall of tidally stripped NS material \citep[e.g.,][] 
{lr07}.

\begin{figure}
\resizebox{\hsize}{!}{\includegraphics[ ]{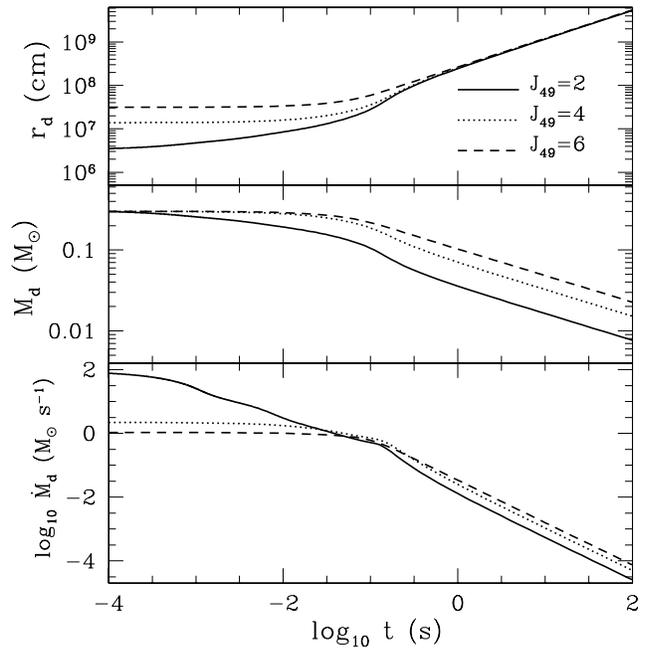}}
\caption{Similar to Fig. \ref{fig:mass1}, but now taking the angular
momentum to be $J_{49}=2$ ({\it solid lines}), $4$ ({\it dotted
lines}), and $6$ ({\it dashed lines}). All solutions take
$M_{d,0}=0.3M_\odot$ with all other variables the same as in
Fig. \ref{fig:mass1}.}
\label{fig:mass2}
\end{figure}

As an additional comparison, we present the effect of varying $J_d$ in
Figure \ref{fig:mass2}. The main trend is that a higher $J_d$ has a
larger initial radius for a given $M_d$, and therefore a longer
viscous time and smaller accretion rate.  The late time behavior is
more sensitive to $J_d$ than the initial $M_d$, as predicted by the
self-similar solutions, but it still does not affect the late time
disk radius (see eq. [\ref{eq:rd_analytic2}]). We do not plot our
results for different $\alpha$ since they are generally consistent
with the analytic scalings above and in Appendix \ref{sec:analytic}.

\subsection{Composition}
\label{sec:composition}
 
\begin{figure}
\resizebox{\hsize}{!}{\includegraphics[ ]{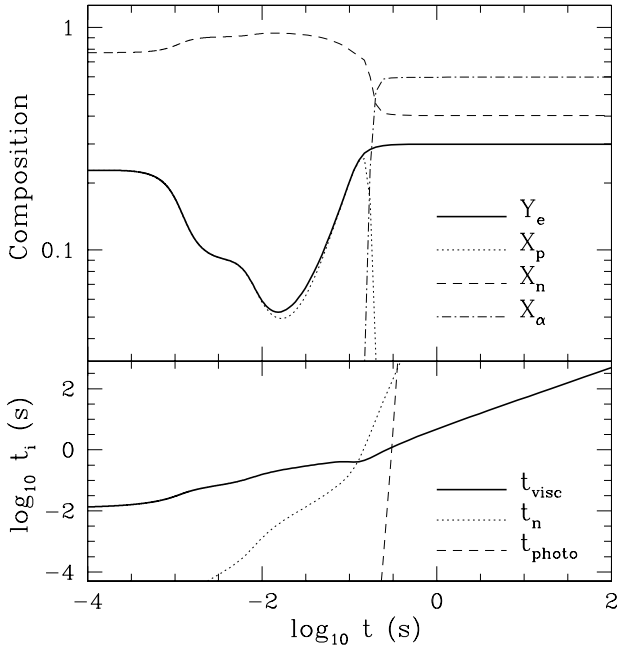}}
\caption{The composition and important reaction
timescales as a function of time, for the $M_{d,0}=0.3M_\odot$ model from
Fig. \ref{fig:mass1}.  In the top panel we plot the electron fraction,
$Y_e$ and the mass fraction of protons, neutrons, and
$\alpha$-particles ({\it see inset key}). In the bottom panel we show
the viscous time, $t_{\rm visc}$ ({\it thick, solid line}), the
neutronization time, $t_n=1/r_{e^-p}$ ({\it dotted line}), and the
$\alpha$-particle photodisintegration time, $t_{\rm photo}$ ({\it
dashed line}).}
\label{fig:composition}
\end{figure}

   The composition of the disk is important for determining the
observational effects of any outflows. To this end, we plot the
composition of our $M_{d,0}=0.3\ M_\odot$, $J_{49}=2$ disk as a function
of time in the upper panel of Figure \ref{fig:composition}.  In the
bottom panel we plot the relevant timescales for setting the
composition, namely the viscous timescale, $t_{\rm visc}$ ({\it solid
line}), the neutronization timescale $t_n=1/r_{e^-p}$ ({\it dotted
line}), and the timescale for $\alpha$-particle photodisintegration,
$t_{\rm photo}$ ({\it dashed line}).  At early times $t_n\ll t_{\rm
visc}$, so that an equilibrium value of $Y_e\simeq0.23$ is reached almost immediately. As the disk leaves the optically thick phase and
becomes thinner, degeneracy pressure plays a larger role. This
enhances neutron production, with a minimum $Y_e\simeq0.05$. As the
neutrino cooling subsides and the disk becomes thick again, $Y_e$
increases. Before $Y_e$ can reach $\simeq0.5$, it freezes-out at a
value of $Y_e\simeq0.3$ once $t_n>t_{\rm visc}$.


Besides the neutron abundance, Figure \ref{fig:composition} also
highlights the production of $\alpha$-particles. Initially, the reactions needed
to convert neutrons and protons to helium as well as
photodisintegration of helium all happen on timescales much shorter
than the disk evolution timescale (as an example, we plot the helium
photodisintegration timescale in the bottom panel of
Fig. \ref{fig:composition}), so that we can estimate the
$\alpha$-particle mass fraction using chemical balance
(eq. [\ref{eq:saha}]). Once the $\alpha$-particle photodisintegration
timescale becomes sufficiently long ($t_{\rm visc}<t_{\rm photo}$),
chemical equilibrium no longer applies and $X_p=0$,
$X_n=1-2Y_e\simeq0.4$, and $X_\alpha=2Y_e\simeq0.6$.

\begin{figure}
\resizebox{\hsize}{!}{\includegraphics[ ]{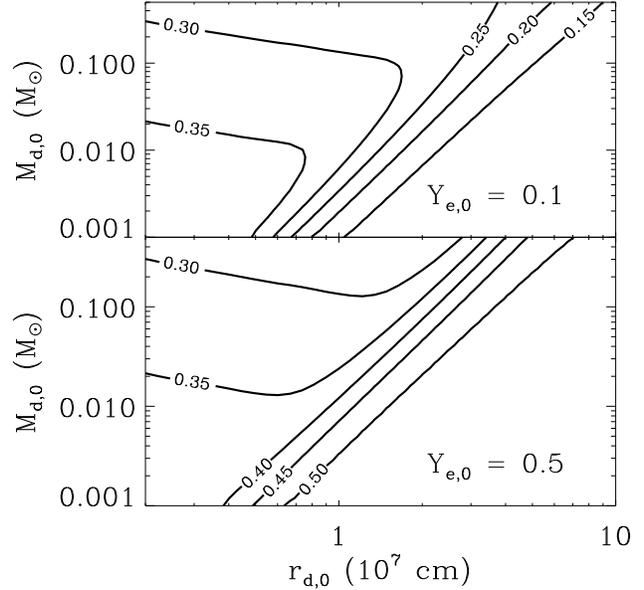}}
\caption{Contours of late-time electron fraction in the expanding disk
as a function of initial disk mass $M_{d,0}$ and radius $r_{d,0}$, for
two different initial compositions.  Relatively compact disks come
into $\beta$-equilibrium and reach an electron fraction independent of
the initial $Y_e$, while low mass, more extended disks retain their
initial composition.  Figure \ref{fig:composition} shows the evolution
of $Y_e$ with time for one particular disk solution.}
\label{fig:Ye}
\end{figure}

Figure \ref{fig:Ye} shows how the late-time, frozen-out value of $Y_e$ in the disk depends
on the initial disk mass $M_{d,0}$ and radius $r_{d,0}$, for two
different initial electron fractions, $Y_{e,0} = 0.1$ and $Y_{e,0} =
0.5$.  The former is relevant for the disks created from NS-NS or
BH-NS mergers (the focus of this paper), while a larger $Y_{e,0}
\simeq 0.5$ is appropriate for disks created during the
accretion-induced collapse of a white-dwarf to a neutron star (e.g., Woosley $\&$ Baron 1992; Dessart et al.~2006).  Figure \ref{fig:Ye} shows that for sufficiently compact disks, the disk
reaches a modestly neutron-rich composition, with $Y_e \simeq
0.3-0.4$, independent of the initial composition.  This is because,
as highlighted in Figure \ref{fig:composition}, the timescale to come
into $\beta$-equilibrium is shorter than the viscous time.  For disks
with a small initial mass and/or a large initial radius (the lower right-hand corner
of each panel), $t_n>t_{\rm visc}$
and the disk retains its initial composition (set by the tidally-disrupted progenitor
and the subsequent dynamical stage of the merger).
Finally, neutrino irradiation of the outer disk by the inner disk can increases the freeze-out electron fraction, but we estimate this changes the freeze-out value of $Y_{e}$ by at most $\sim 20\%$.\footnote{Our calculations employ the pair-capture cooling prescription of DiMatteo et al.~(2002), which assume $Y_{e} = 0.5$ and ultra-relativistic electrons; we find, however, that including the effects of degeneracy and arbitrary electron energies on the cooling changes the asymptotic electron fraction by at most a few percent.}

\section{Disk Winds}
\label{sec:winds}

Having described the evolution of the accretion disk as a function of
time, we now discuss the properties of outflows from these
hyper-accreting disks. Winds driven from deep within the BH potential
well could produce relativistic jets and power late-time central
engine activity.  Outflows driven from larger radii dominate the
system's mass loss and may power supernova-like optical transients
through the decay of radioactive isotopes that are synthesized in the
wind (Li \& Paczy{\'n}ski 1998; Kulkarni 2005). In both cases, the
mass loss rate and nuclear composition are critical for determining
the observable signature.

The type and character of the outflow depends on the disk's
thermodynamic state and changes as it passes through the different
stages of evolution described in the previous section. In \S
\ref{thindiskwinds} we discuss early times when winds are due to
neutrino irradiation of the thin, efficiently neutrino-cooled portions
of the disk.  We then consider thermally driven winds during thick,
radiatively-inefficient accretion in \S \ref{thickdiskwinds}. This
dominates the mass loss at late times and blows away most of the
remaining disk.  In \S\ref{nuclearcomposition} we summarize the
nuclear composition of the outflows during each phase. We predict an
ejected $^{56}$Ni mass of at most $\sim 10^{-3} M_{\sun}$ (\S
\ref{sec:ni}). Its decay may power transient emission detectable
following some short GRBs.

\subsection{Neutrino-Heated Thin-Disk Winds}
\label{thindiskwinds}

A wind with a mass loss rate $\dot{M}_w$ driven from a thin disk at
radius $r$ must absorb a net power greater than $\dot{E}_{b} =
GM\dot{M}_w/2r$ to become unbound from the central BH.  In principle,
$\dot{E}_{b}$ may be supplied by dissipation of the turbulence that
produces the accretion shear stresses.  ``Viscous'' heating of this
kind only efficiently drives an outflow if a substantial fraction of
the accretion power is dissipated in the disk's upper atmosphere,
where the cooling timescale is long compared to the wind's outward
advection timescale.  However, local radiation MHD simulations to date
suggest that very little energy dissipation occurs in the corona
(e.g., Krolik et al.~2007).  Instead, heating in the atmosphere above
a thin, neutrino-cooled disk is likely dominated by neutrino
irradiation.  We therefore focus on the neutrino-driven mass loss
rate, which sets a \emph{minimum} $\dot{M}_w$, and which can be
reliably estimated.  Neutrino-driven outflows from hyper-accreting disks have also been studied by Daigne $\&$ Mochkovitch (2002), Levinson (2006), Metzger et al.~(2008a), and Barzilay $\&$ Levinson (2008); Dessart et al.~(2008b) calculate the neutrino-driven mass loss from the central NS following a NS-NS merger under the assumption that collapse to a BH is not prompt.

   The neutrino-driven mass loss rate is calculated by equating $\dot{E}_b$ to
the total neutrino heating rate in the disk's atmosphere. For the
radii and entropies that characterize the winds, heating via electron
neutrino absorption on baryons ($p+\bar{\nu}_{e}\rightarrow n+e^{+}$
and $n+\nu_{e}\rightarrow p+e^{-}$) dominates other forms of neutrino
heating (e.g., $\nu-\nu$ annihilation and $\nu-e^{-}$ scattering; see
Qian $\&$ Woosley 1996; hereafter QW96).  Since the neutrino
absorption cross section,
$\sigma_{\nu N}\simeq 5\times 10^{-44}\langle\epsilon_{\nu}^{2}\rangle$ MeV$^{-2}$ cm$^{2}$,
increases with neutrino energy, neutrinos radiated from near the inner
radius $r_{*}$ dominate. Assuming that the $\nu_{e}$ and $\bar{\nu}_{e}$
luminosities and spectra are approximately equal and can be approximated
as originating from a point source at small radii, the neutrino heating rate
through a surface density $\Sigma$ at radius $r$ is
\be
	q_{\nu}^{+} = \frac{L_{\nu}\sigma_{\nu N}\Sigma}{4\pi m_N r^{2}} 
	\simeq 2\times 10^{39}L_{52}\langle\epsilon_{10}^{2}\rangle
	\Sigma_{18}r_{6}^{-2}{\rm \,ergs\,\,s^{-1}\,\,cm^{-2}},
	\nonumber
	\\
	\label{eq:qnuplus}
\ee
where $r = 10^{6}r_{6}$ cm, $L_{\nu} = 10^{52}L_{52}$ ergs s$^{-1}$,
$\langle\epsilon_{\nu}^{2}\rangle$ = 100$\langle\epsilon_{10}^{2}\rangle$ MeV$^{2}$,
and $\Sigma = \Sigma_{18}10^{18}$ g cm$^{-2}$.  This expression assumes that the absorbing
layer is optically thin, i.e., that
$\tau_{\nu} \equiv \Sigma\sigma_{\nu N}/m_{N} \simeq 3\Sigma_{18}\langle\epsilon_{10}^{2}\rangle < 1$.

   First, consider neutrino heating in comparison to viscous heating in the midplane.  This ratio
is largest when the disk is marginally optically thick ($\tau_{\nu} \simeq 1$),
peaking at a value of
\bea
	\left.\frac{q_{\nu}^{+}}{q_{\rm visc}}\right|_{\tau_{\nu} \simeq 1} \simeq \nonumber \eea \bea 0.5\left(\frac{\epsilon}{0.1}\right)\left(\frac{f}{1.6}\right)\left(\frac{A}{3.6}\right)^{3/5}\langle\epsilon_{10}^{2}\rangle^{2/5}J_{49}^{2/5}M_{3}^{-6/5},
\label{eq:heating_ratio}
\eea where $\epsilon \equiv L_{\nu}/\dot{M}_{d}c^{2}$ is the disk's
radiative efficiency.  Thus, although we neglected neutrino heating in
\S \ref{sec:results}, it may become somewhat important when $\tau_\nu
\sim 1$ and should be included in a more detailed calculation.

We now consider a wind that emerges from the disk in the
$z$-direction, parallel to the rotation axis. Away from the disk
midplane, neutrino heating dominates over viscous heating, balancing
cooling ($q_{\nu}^{+} = q_{\nu}^{-}$) at a slightly lower temperature,
$ T_{\nu} \simeq
3.3L_{52}^{1/6}\langle\epsilon_{10}^{2}\rangle^{1/6}r_{6}^{-1/3}{\rm
MeV}$.  Moving further out in the hydrostatic atmosphere, the
temperature slowly decreases below $T_{\nu}$.  Due to the strong
temperature dependence of the pair capture cooling rate ($q_{\nu}^{-}
\propto T^{6}$), a ``gain region'' of net neutrino heating (i.e.,
$q_{\nu}^{+} > q_{\nu}^{-}$) develops above a height $z_{\rm
gain}$. This net heating drives an outflow.

The thermal power deposited in the upper disk atmosphere
$\dot{E}_{\nu}$ is the specific heating rate $q^{+}_{\nu}/\Sigma$
(eq.~[\ref{eq:qnuplus}]) multiplied by the mass of the atmosphere in
the gain region $M_{\rm gain} \simeq 2\pi H(z_{\rm
gain})r^{2}\rho(z_{\rm gain})$, where $H(z_{\rm gain})$ is the scale
height near the base of the gain region.  Although the midplane of a
neutrino-cooled disk is generally dominated by nonrelativistic gas
pressure (see Fig. \ref{fig:pressure}), the gain region has a
sufficiently low density that it is instead dominated by radiation
pressure $P_{\rm rad} = (11/12)a T^{4}$. Its scale height is $H(z_{\rm
gain}) \simeq (P_{\rm rad}/\rho g_{\rm z})|_{\rm z_{\rm gain}}$, where
$g_{\rm z}$ is the gravitational acceleration in the
$z$-direction. Since $H(z_{\rm gain})$ is less than the midplane scale
height $H$, $z_{\rm gain} \simeq$ H and $g_{\rm z} \simeq
GMH/r^{3}$. The atmosphere in the gain region is roughly isothermal so
we set $T(z_{\rm gain}) \approx T_{\nu}$.  By combining these
estimates and equating $\dot{E}_{\nu}$ with $\dot{E}_{b}$ we find that
the neutrino-driven mass loss rate from a thin disk is \be
\dot{M}_{\nu}|_{S^{a} \gg S_{\rm N}} \approx 10^{-6}L_{52}^{5/3}
\langle\epsilon_{10}^{2}\rangle^{5/3}r_{6}^{5/3}M_{3}^{-2}(H/r)^{-1}M_{\sun}{\rm s^{-1}},
	\label{eq:mdot_nu}
\ee
analogous to that derived by QW96 for proto-neutron star winds.  The assumption that the atmosphere is
radiation dominated is only valid if the asymptotic entropy in relativistic particles $S^{a}$ exceeds that in
nonrelativistic nucleons $S_{\rm N} \simeq 6 + {\rm ln}(T_{\rm MeV}^{3/2}/\rho_{10}) k_{B}$ baryon$^{-1}$,
where $T = T_{\rm MeV}$ MeV.  By dividing the energy gained by a nucleon in the wind
$\simeq GMm_{N}/2r$ by the gain region temperature $T(z_{\rm gain})$, we estimate
\be
	S^{a} \simeq 60 L_{52}^{-1/6}\langle\epsilon_{10}^{2}\rangle^{-1/6}r_{6}^{-2/3}M_{3}\,\,k_{\rm B}{\rm\,baryon^{-1}}
	\label{eq:sa}
\ee
as the asymptotic wind entropy.

Although equation (\ref{eq:mdot_nu}) does not strictly hold when $S^{a}\sim S_{\rm N}$, QW96 show that
$\dot{M}_{\nu}$ scales the same way with $L_{\nu}$, $\langle\epsilon_{\nu}^{2}\rangle$, $M$, and $r$,
but with a larger normalization of
\be
	\dot{M}_{\nu}|_{S^{a} \sim S_{\rm N}} \approx 10^{-5}L_{52}^{5/3}
	\langle\epsilon_{10}^{2}\rangle^{5/3}r_{6}^{5/3}M_{3}^{-2}(H/r)^{-1}M_{\sun}{\rm s^{-1}}.
	\label{eq:mdot_nu2}
\ee The mass loss rate is higher for low entropy winds because
neutrino heating peaks further off the disk surface, which reduces the
binding energy and gravitational acceleration of matter in the gain
region.  Using the numerical disk wind calculations described in
Metzger et al. (2008b; hereafter M08b) we have verified that equation
(\ref{eq:mdot_nu2}) holds to within a factor $\simeq 2$ when $S^{a}
\sim S_{\rm N}$.

In deriving equations (\ref{eq:mdot_nu}) and (\ref{eq:mdot_nu2}), we
have implicitly assumed that the timescale for neutrinos to heat matter in the gain region
$t_{\rm heat} \equiv (U_{\rm th}\Sigma/\rho q_{\nu}^{+})|_{z_{\rm gain}}$, where
$U_{\rm th} \simeq 3P_{\rm rad}$ is the thermal energy density, is short compared to $t_{\rm visc}$,
the timescale over which the disk properties appreciably change.  Equating $S^{a}$ (eq.~[\ref{eq:sa}])
to the entropy in relativistic particles $\propto T^{3}/\rho$, we find that
\be \rho(z_{\rm gain}) \simeq 10^{8} r_{6}^{-1/3}L_{52}^{2/3}\langle\epsilon_{10}^{2}\rangle^{2/3}M_{3}^{-1} {\rm g\,\,cm^{-3}}. 
\label{eq:rho_gain}
\ee
Then, using equations (\ref{eq:qnuplus}) and (\ref{eq:rho_gain}), we have
that\footnote{Equation (\ref{eq:theat}) is also approximately equal to the outward advection
timescale of the wind in the heating region.} 
\be
	t_{\rm heat} \simeq \left.\frac{3P_{\rm rad}}{\rho(q^{+}_{\nu}/\Sigma)}\right|_{z_{\rm gain}}
	\simeq 0.1\,{\rm s}\,\,L_{52}^{-1}r_{6}\langle\epsilon_{10}^{2}\rangle^{-1}M_{3}
	\label{eq:theat}
\ee  
For most of the disk solutions considered in this paper, we find that
$t_{\rm heat} \lesssim t_{\rm visc}$ during the thin disk phase; thus, equations (\ref{eq:mdot_nu})
and (\ref{eq:mdot_nu2}) are reasonably applicable near $r_{d}$.

\begin{figure}
\resizebox{\hsize}{!}{\includegraphics[ ]{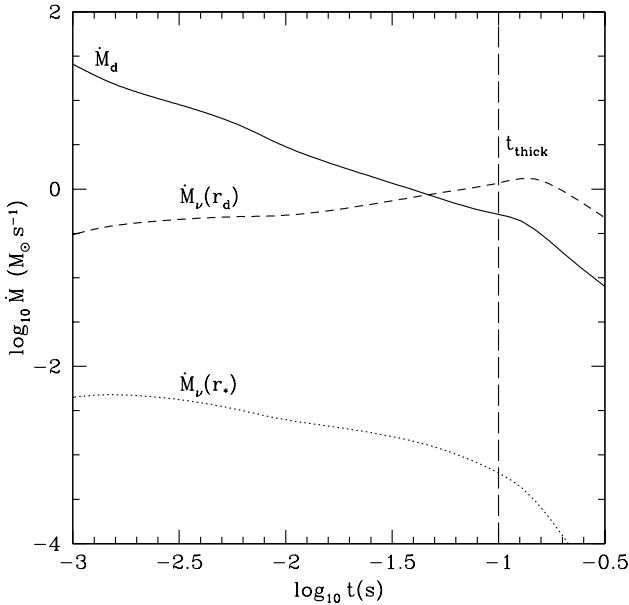}}
\caption{The accretion rate $\dot{M}_{d}$ ({\it solid line}) and
neutrino-driven mass loss rates $\dot{M}_\nu$ for our $J_{49} = 2$ and
$M_{d,0} = 0.3 M_{\sun}$ model, focusing on the phase of thin,
efficiently neutrino-cooled accretion. The neutrino-driven mass loss
rate $\dot{M}_{\nu}$ (interpolated between eqs.~[\ref{eq:mdot_nu}] and
[\ref{eq:mdot_nu2}]) is shown at the inner disk radius ($r_{*} =
10^{6}$ cm; {\it dotted line}) and at the outer disk radius (near
$r_{d}$, {\it short-dashed line}).  The disk is advective to the right
of the vertical line (eq. [\ref{eq:tthick}]), at which point the
mass-loss will no longer be dominated by neutrino irradiation.}

\label{fig:mdot}
\end{figure}

   Figure \ref{fig:mdot} compares the accretion rate $\dot{M}_{d}$
({\it solid line}) with the neutrino-driven mass loss rate
$\dot{M}_\nu$. In order to determine $L_{\nu}$ and
$\langle\epsilon_{\nu}^{2}\rangle$, we calculated steady-state disk
models \cite[e.g.,][]{dim02} with the accretion rate set at each time
according to our ring model with $J_{49} = 2$ and $M_d=0.3
M_{\sun}$.  We plot the neutrino-driven mass loss rate $\dot{M}_{\nu}$
(eqs.~[\ref{eq:mdot_nu}] and [\ref{eq:mdot_nu2}]) at small ({\it
dotted line}) and large ({\it short-dashed line}) radii. This shows
that the mass loss is dominated by large radii where the majority of
the mass lies, as expected since $\dot{M}_{\nu} \propto r^{5/3}$.  The vertical dot-dashed line marks where the
disk transitions to being thick (eq. [\ref{eq:tthick}]),
after which neutrino-heating no longer dominates the wind mass loss.

Outflows that are launched from small radii, near $r_{*}$, have the
greatest potential to produce relativistic jets and to power high
energy emission. But as we now argue, these neutrino-driven winds are
too massive to become highly relativistic. Our calculation above
focused on purely thermal, neutrino-driven winds, which accelerate
matter to only a fraction of the escape speed (and thus are mildly relativistic). However, in the presence of a strong, large scale
open poloidal magnetic field, a more powerful, magnetically-driven
outflow is possible.  Magnetocentrifugal support in the wind's
hydrostatic atmosphere may further enhance mass loss (e.g., Levinson 2006), but equation (\ref{eq:mdot_nu}) still represents the {\it minimum} mass loading on
field lines which thread a neutrino-cooled disk.  Figure
\ref{fig:mdot} shows that $\dot{M}_{\nu}(r_{*}) \sim
10^{-4}-10^{-2}M_{\sun}$ s$^{-1}$ during the thin disk phase.  The
luminosities of the prompt emission and late-time X-ray flares from
short GRBs, however, do not typically exceed $L_\gamma \sim 10^{50}$
erg s$^{-1}$ (and are often much lower; Nakar 2007).  Thus, even
assuming a modest radiative efficiency for the outflow of
$\epsilon_{w} \sim 0.1$, the Lorentz factor $\Gamma$ of a
neutrino-heated disk wind must obey $\Gamma \simeq
L_\gamma/[\epsilon_{w}\dot{M}_{\nu}(r_{*})c^{2}] \lesssim 5$, which is
inconsistent with existing compactness constraints on short GRBs
(Nakar 2007).  A more likely source for the relativistic outflows that
power short GRBs and their late-time flares are nearly baryon-free
field lines which thread the BH's event horizon (e.g., McKinney 2005).
In addition, in \S \ref{thickdiskwinds} we argue that when the disk becomes
advection dominated and neutrino irradiation effectively ceases, jet
production may be more likely.

\subsection{Radiatively-Inefficient Thick Disk Winds}
\label{thickdiskwinds}

At late times ($t \sim t_{\rm thick}$; eq.~[\ref{eq:tthick}]) the disk
transitions from thin and neutrino-cooled to being advective.  At this
point a neutrino-driven outflow is unlikely to dominate the mass loss,
in part because the neutrino luminosity precipitously drops
(Fig. \ref{fig:energy1} \& \ref{fig:energy2}).  In addition, because
RIAFs possess a positive Bernoulli parameter, a powerful
viscously-driven outflow is likely (Blandford $\&$ Begelman 1999;
Stone $\&$ Pringle 2001; Proga $\&$ Begelman 2003).

In \S \ref{sec:gross} we showed that the disk becomes radiatively
inefficient external to an ``ignition radius'' $r_{\rm ign} \propto
\dot{M}_{d}^{6/5}$ (eq.~[\ref{eq:rign}]). The outer disk, near
$r_{d}$, thickens first (when $r_{d} \sim r_{\rm ign}$ at $t \sim
t_{\rm thick}$) and radiatively inefficient conditions move inwards as
$\dot{M}_{d}$ decreases. In the simplest picture, one might expect
that the innermost radii become an RIAF only once $\dot{M}_{d}$ drops
from its value at $t \sim t_{\rm thick}$ by an additional factor $\sim
(r_{*}/r_{d})^{5/6}$. In fact, the {\it entire} disk probably become
radiatively inefficient on a timescale similar to $t_{\rm thick}$ if
the accretion rate which reaches small radii abruptly decreases once
the outer disk thickens (Fig.~\ref{fig:wind}).  Hence, at a time $t_{\rm thick}$, a
significant portion of the accreting matter may be redirected into an
outflow, with only a fraction $\sim (r_{*}/r_{d})$ reaching small
radii and accreting onto the BH (Stone $\&$ Pringle 2001).

X-ray binaries typically produce radio jets upon transitioning from
their ``high-soft'' (radiatively efficient) to ``low-hard''
(radiatively inefficient) states (e.g., Remillard $\&$ McClintock
2006).  In analogy, once the inner disk becomes an RIAF, conditions
seem to favor the production of relativistic jets (see also Lazzati et al.~2008).\footnote{This is in
stark contrast to jets powered by neutrino annihilation along the
polar axis, which require a {\it high} radiative efficiency.} 

Even if only a fraction $(r_{*}/r_{d})$ of the mass remaining when the
disk thickens actually reaches the origin, the total energy supply
available would be \bea E_{\rm jet}&\equiv&\epsilon_{\rm
jet}M_d(t_{\rm thick})c^{2}\left(\frac{r_{*}}{r_d(t_{\rm
thick})}\right) \nonumber \eea \bea \simeq 3\times
10^{50}\left(\frac{\epsilon_{\rm jet}}{0.1}\right)
\left(\frac{r_{*}}{10^{6}{\rm
cm}}\right)\alpha_{0.1}^{6/17}M_{3}^{-4/17}
\left(\frac{J_{49}}{2}\right)^{8/17}{\rm ergs}, \nonumber \\
	\label{eq:ejet}
\eea where $\epsilon_{\rm jet}$ is the fraction of the accretion energy
used to power a jet and we have estimated $M_d(t_{\rm thick})$ and
$r_{d}(t_{\rm thick})$ using the self-similar thick disk solutions
(eqs.~[\ref{eq:md_analytic2}] and [\ref{eq:rd_analytic2}],
respectively).  Equation (\ref{eq:ejet}) shows that the accretion
energy available from near $r_{*}$ following the RIAF transition is
more than sufficient to power the late-time X-ray flares observed
following some short GRBs.  If this is the case, $t_{\rm thick}$ sets
a characteristic timescale for late-time central engine activity.  If
$\alpha\lesssim 10^{-3}$, $t_{\rm thick}$ may be large enough to
explain the $\sim 30\ {\rm s}$ delay until flaring observed for some
short GRBs (e.g., Berger et al.~2005; Villasenor et
al.~2005). However, very late time energy injection, such as the {\it
Chandra} flare observed two weeks after GRB050709 (Fox et al.~2005),
appears to require an alternative explanation. In addition, given
observational evidence for $\alpha \sim 0.1$ in a number of
environments (King et al. 2007), it may be more natural to associate
$E_{\rm jet}$ and $t_{\rm thick}$ with the energy and duration,
respectively, of the short GRB itself, rather than the late-time
central engine activity (see \S \ref{sec:theend}).

\subsection{Outflow Nuclear Composition}
\label{nuclearcomposition}

The outflow nuclear composition has important consequences for the
observable signature of compact object mergers. Nonrelativistic
outflows are sufficiently dense to synthesize heavy isotopes (Pruet et al.~2004; Surman et al.~2006), which
may power transient emission via radioactive decay.  The isotopic
yield depends on the speed, thermodynamic properties, and the asymptotic electron fraction $Y_{e}^{a}$ in the
outflow.\footnote{The {\it asymptotic} electron fraction is germane
because heavy nuclei primarily form after freeze-out from
$\beta$-equilibrium.}  Although relativistic winds from the inner disk
are unlikely to synthesize anything heavier than He (Lemoine 2002;
Beloborodov 2003a), $Y_{e}^{a}$ is important in this
case as well. A neutron-rich outflow may alter the jet's dynamics and
the prompt and afterglow emission from that of the standard GRB
fireball model (e.g., Derishev et al.~1999; Beloborodov 2003b; Rossi
et al.~2006).

\begin{figure}
\resizebox{\hsize}{!}{\includegraphics[ ]{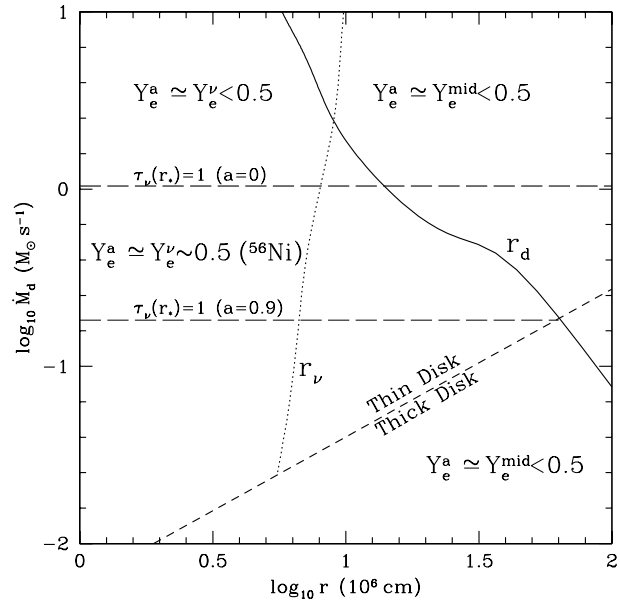}}
\caption{Asymptotic electron fraction $Y_e^a$ for disk winds as a
function of the wind launching radius $r$ and accretion rate
$\dot{M}_{d}$ (for $\alpha = 0.1$ and $M = 3M_{\sun}$). The solid line
indicates the location of the ring radius $r_{d}$ for our fiducial
solution with $M_{d,0} = 0.3 M_{\sun}$ and $J_{49} = 2$.  The short
dashed line is the ``ignition'' radius $r_{\rm ign}$
(eq.~[\ref{eq:rign}]).  Exterior to this (marked ``Thick Disk'') the
disk is advective with a viscously driven wind of composition
$Y_e^a\simeq Y_e^{\rm mid}<0.5$.  Interior to $r_{\rm ign}$ (marked
``Thin Disk'') a neutrino-driven wind occurs. The dotted line shows
$r=r_\nu$ with $Q=2$ (eq. [\ref{eq:rnu}]) and determines where the
neutrino absorptions necessary to unbind matter alter the wind
composition, so that $Y_e^a\simeq Y_e^{\rm mid}<0.5$ ($Y_e^a\simeq
Y_e^\nu$) exterior (interior) to $r_\nu$.  The $\dot{M}_{d}$ above
which $\tau(r_{*})>1$ is plotted for BH spins of $a = 0$ and $a =
0.9$.  Above this line, the $\bar{\nu}_{e}$ and $\nu_{e}$ spectra
differ and $Y_{e}^{\nu} <0.5$, while below this their spectra are
similar and $Y_{e}^{\nu} \simeq 0.5$.  In the region where $r < r_{\rm
ign}$, $\tau_{\nu}(r_{*}) < 1$, and $r<r_\nu$ (i.e., the middle/lower left-hand trapezoid), $Y_{e}^{a} \simeq Y_{e}^{\nu} \sim 0.5$; these
conditions are favorable for $^{56}$Ni production (see
$\S\ref{sec:ni}$).}
\label{fig:mdot_radius}
\end{figure}

Figure \ref{fig:mdot_radius} delineates different regimes of outflow properties and
composition (as given by $Y_e^a$) as a function of the wind launching
radius $r$ and accretion rate $\dot{M}_{d}$. We fix $\alpha = 0.1$ and
$M = 3M_{\sun}$. The time-dependent evolution of the ring radius $r_{d}$
is shown for a solution with $J_{49} = 2$ and $M_{d,0} = 0.3M_{\sun}$ ({\it solid line}).
At each time a given steady-state disk profile can be read off of this plot as a horizontal line
that extends from the far left and ends on $r_d$.
Therefore, outflows from radii {\it interior} to $r_{d}$ contribute to the disk's total nucleosynthetic yield.

The ignition radius $r_{\rm ign}$ (eq.~[\ref{eq:rign}]) is shown in
Figure \ref{fig:mdot_radius} with a short dashed line. For $r \gtrsim
r_{\rm ign}$ the disk is an RIAF and marked in the figure as ``Thick
Disk.'' In this case, a viscously driven outflow dominates (\S
\ref{thickdiskwinds}). Since outflows from \mbox{RIAFs} escape the
disk in roughly the accretion timescale, these winds retain the
midplane electron fraction (M08b), so that $Y_e^a\simeq Y_{e}^{\rm
mid}\ll 0.5$ (because the disk itself freezes-out neutron-rich, as summarized in \S \ref{sec:composition} and
Fig. \ref{fig:Ye}).

For $r \lesssim r_{\rm ign}$, the disk is efficiently neutrino-cooled
and marked in Figure \ref{fig:mdot_radius} as ``Thin Disk.'' The
absorption of neutrinos, which heats the outflow and unbinds it from
the BH may also alter its nucleonic composition. This drives
$Y_e^a$ to a value set by the neutrino radiation field $Y_e^\nu$,
which in general is different from $Y_{e}^{\rm mid}$.  A simple
criterion was discussed by M08b for determining when $Y_e^a\simeq
Y_e^\nu$.  A typical nucleon in the accretion disk at radius $r$ must
absorb an energy $\simeq GMm_{N}/2r$ to become unbound from the BH, so that $N_{\nu} \simeq GMm_{N}/2r\langle\epsilon_{\nu}\rangle$
neutrinos must be absorbed per nucleon. If we take $N_{\nu} > Q \sim
2-3$, then a typical nucleon has changed its identity ($p\rightarrow
n$ or $n\rightarrow p$) at least several times.

This implies that all purely neutrino-driven outflows from radii
smaller than \be r_{\nu} \equiv
\frac{GMm_{p}}{2Q\langle\epsilon_{\nu}\rangle} \simeq
10^{7}M_{3}\langle\epsilon_{10}\rangle^{-1}(Q/2)^{-1}{\rm\,cm},
	\label{eq:rnu}
\ee where $\langle\epsilon_{\nu}\rangle \equiv
10\langle\epsilon_{10}\rangle\ {\rm MeV}$, achieve $Y_{e}^{a} \simeq
Y_{e}^{\nu}$, independent of the disk's midplane composition.

We plot $r_\nu$ with $Q = 2$ as a dotted line in Figure \ref{fig:mdot_radius}, where $\langle\epsilon_{\nu}\rangle$ is calculated from $\dot{M}_{d}$ using our steady-state disk solutions (see $\S \ref{thindiskwinds}$).  For $r \lesssim r_{\nu}$, any neutrino-driven outflow enters equilibrium with the neutrino radiation field (i.e., $Y_{e}^{a} \simeq Y_{e}^{\nu}$).  For $r \gtrsim r_{\nu}$ the outflow approximately retains the midplane electron fraction (i.e., $Y_{e}^{a} \simeq Y_{e}^{\rm mid}$).

Although we have established the conditions under which $Y_{e}^{a}$ is
determined by neutrino absorptions, we must now address what sets
$Y_{e}^{\nu}$ itself.  If the rate of neutrino absorptions exceeds the
rate of degenerate pair captures before the wind falls out of
$\beta$-equilibrium, $Y_e^\nu$ is \be Y_e^\nu \equiv \left(1+\frac{L_{\bar{\nu}_e}}{L_{\nu_e}}
\frac{\langle\epsilon_{\bar{\nu}_e}\rangle-2\Delta +
1.2\Delta^2/\langle\epsilon_{\bar{\nu}_e}\rangle}
{\langle\epsilon_{\nu_e}\rangle+2\Delta+1.2\Delta^2/\langle\epsilon_{\nu_e}\rangle}
\right)^{-1},
	\label{eq:yeanu}
\ee where $\Delta = 1.293$ MeV is the neutron-proton mass difference,
and $L_{\nu_{e}}/L_{\bar{\nu}_{e}}$ and
$\langle\epsilon_{\nu_{e}}\rangle$/$\langle\epsilon_{\bar{\nu}_{e}}\rangle$
are the mean $\nu_{e}$/$\bar{\nu}_{e}$ luminosities and energies,
respectively, from a centrally-concentrated source (Qian et al.~1993;
QW96). Equation (\ref{eq:yeanu}) demonstrates that the $\nu_e$ and
$\bar{\nu}_e$ spectra are crucial for setting $Y_e^\nu$.
 
Since the disk's luminosity and temperature peak at just a few
$r_{g}$, $Y_{e}^{\nu}$ is primarily determined by conditions at small
radii. At early times, the accretion disk may be optically thick near
$r_{*}$ and so the $\nu_{e}$ and $\bar{\nu}_{e}$ spectra depend on the
temperatures at $\nu_{e}$ and $\bar{\nu}_{e}$ neutrinospheres,
respectively.  Since there are more neutrons than protons in the disk,
the optical depth to $\nu_{e}$ through the disk is higher than to
$\bar{\nu}_{e}$; thus, the temperature at the $\bar{\nu}_{e}$
neutrinosphere is higher than at the $\nu_{e}$ neutrinosphere.  This
implies $L_{\bar{\nu}_{e}} \gg L_{\nu_{e}}$,
$\langle\epsilon_{\bar{\nu}_{e}}\rangle \gg
\langle\epsilon_{\nu_{e}}\rangle$, and thus $Y_{e}^{\nu} \ll 0.5$.
Using 3-dimensional calculations of the merger of NSs with zero spin, Rosswog $\&$ Liebend{\"o}rfer (2003) find that at $\sim 15$ ms following merger,
$L_{\bar{\nu}_{e}} \simeq 3.5L_{\nu_{e}}$,
$\langle\epsilon_{\nu_{e}}\rangle \simeq 9$ MeV, and
$\langle\epsilon_{\bar{\nu}_{e}}\rangle \simeq$ 15 MeV, which implies
$Y_{e}^{\nu} \simeq 0.21$, consistent with our arguments (see also
Surman et al. 2008).  We conclude that when the disk is optically-thick near $r_{*}$, a
neutron-rich outflow is again the most likely outcome.  The critical accretion rate at which $\tau_{\nu}(r_{*})=1$ is shown in Figure \ref{fig:mdot_radius} with a long dashed line for both $a = 0$ and $a=0.9$.

Once the disk becomes optically thin near $r_{*}$, the
difference between the $\nu_{e}$ and $\bar{\nu}_{e}$ spectra is much
less pronounced.  This occurs because (1) the neutrinos and
antineutrinos originate from regions with the same temperature; (2)
any net lepton flux out of the disk must remain modest (i.e.,
$L_{\nu_{e}}/\langle\epsilon_{\nu_{e}}\rangle \simeq
L_{\bar{\nu}_{e}}/\langle\epsilon_{\bar{\nu}_{e}}\rangle$); and (3)
the difference between the $e^{-}$ and $e^{+}$ capture cross sections
for $kT \gg \Delta-m_{e}c^{2}$ is small. Taking
$\langle\epsilon_{\nu_{e}}\rangle\sim\langle\epsilon_{\bar{\nu}_{e}}\rangle
\gg \Delta$, equation (\ref{eq:yeanu}) shows that $Y_{e}^{\nu}\simeq
0.5$, a value in the range required to produce $^{56}$Ni (which we
discuss further in \S \ref{sec:ni}).  Indeed, M08b used the
steady-state, optically-thin $\alpha$-disk calculations of Chen \&
Beloborodov (2007; hereafter CB07) to calculate the neutrino radiation
fields carefully, and showed that $Y_{e}^{\nu} \gtrsim 0.5$ over the
majority of the disk (see their Fig.~1).  Although the precise spectra
extracted from an $\alpha$-disk calculation should be taken with
caution, the conclusion that the $\nu_{e}$ and $\bar{\nu}_{e}$ spectra
are similar for optically thin accretion (and $Y_{e}^{\nu} \simeq
0.5$) is probably robust.

Figure \ref{fig:mdot_radius} illustrates that under most conditions
the outflows from hyper-accreting disks are neutron-rich.
Neutron-rich material ejected during the initial dynamical phase of
compact object mergers has long been considered a promising source for
producing Galactic $r$-process elements, whose precise astrophysical
origin remains uncertain (Lattimer $\&$ Schramm 1974; see, however,
Qian 2000).  In addition, Surman et al.~(2008) find that winds driven
from the remnant accretion disk at early times (when it is optically
thick; upper left quadrant of Fig. \ref{fig:mdot_radius}) are
sufficiently neutron-rich to produce successful $r$-process.  The
outflows driven from the advective disk at late times, however, are
unlikely to produce r-process elements, given their modest entropies
and electron fractions of $Y_e \gtrsim 0.3$
(Figs. \ref{fig:composition} and \ref{fig:Ye}).  Instead, this modest
$Y_e$ material will be synthesized to form intermediate mass neutron
rich isotopes (Hartmann et al. 1985).

\subsection{$^{56}$Ni Production and Optical Transients}
\label{sec:ni}

As summarized in Figure \ref{fig:mdot_radius}, most of the material in
the outflow driven from a hyper-accreting disk will be neutron-rich.
Nonrelativistic neutron-rich ejecta are difficult to detect because
isotopes synthesized from low $Y_e$ material are themselves very
neutron-rich and typically possess very short half-lives, on the order
of seconds (e.g., Freiburghaus et al.~1999).  Thus, most of the
radioactive energy is released at high optical depths and suffers
severe adiabatic losses before the photons can diffusively escape.  By
contrast, ejecta with $Y_{e}^{a} \simeq 0.5$ are easier to detect
because they can produce a significant quantity of $^{56}$Ni (Hartmann
et al.~1985), an isotope better suited to powering observable emission
because its half-life $\simeq 6$ days is comparable to the timescale
on which the outflow becomes optically thin.  From Figure
\ref{fig:mdot_radius} we see that outflows in a modest range of
parameter space (middle/lower-left trapezoid) are capable of synthesizing
$^{56}$Ni.  One caveat to this conclusion is that it only applies if
the winds are primarily neutrino driven.  If the outflow is instead
magnetocentrifugally driven by a moderately strong open poloidal
magnetic field (e.g., Levinson 2006; Xie et al.~2007), then $Y_{e}^{a} \ll 0.5$ can result, even if
$Y_{e}^{\nu} \simeq 0.5$ (M08b).  In what follows we assume that the
wind's are primarily neutrino driven.

\begin{figure}
\resizebox{\hsize}{!}{\includegraphics[ ]{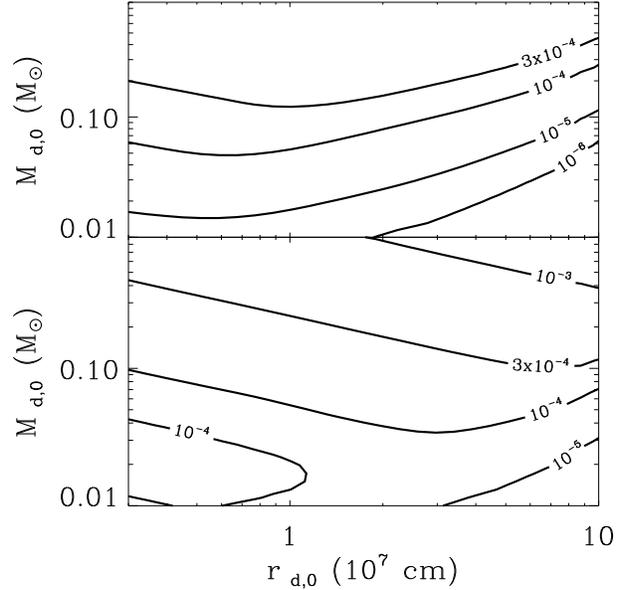}}
\caption{Contours of total $^{56}$Ni mass $M_{\rm Ni} \equiv (X_{\rm
Ni}/0.4)M_{Y_{e} = 0.5}$ (in units of $M_{\sun}$) produced in the
neutrino-driven outflows as a function of the initial disk mass
$M_{d,0}$ and initial ring radius $r_{d,0}$, where $M_{Y_{e} = 0.5}$
is the total mass loss with $Y_{e}^{a} \simeq 0.5$ (based on the
arguments in Fig. \ref{fig:mdot_radius}) and $X_{\rm Ni}$ is the
average $^{56}$Ni mass fraction synthesized in the wind.  The upper
and lower panels correspond to non-rotating ($a = 0$) and rapidly
spinning ($a = 0.9$) BHs, respectively.}
\label{fig:Nimass}
\end{figure}

Under this assumption, Figure \ref{fig:Nimass} shows the total
$^{56}$Ni mass, $M_{\rm Ni}= (X_{\rm Ni}/0.4)M_{Y_{e}=0.5}$, produced
in outflows from hyper-accreting disks as a function of the disk's
initial mass $M_{d,0}$ and radius $r_{d,0}$, where $M_{Y_{e} = 0.5}$
is the total mass loss with $Y_{e}^{a} \simeq 0.5$ and $X_{\rm Ni}$ is
the average $^{56}$Ni mass fraction synthesized in the wind.  We
calculate $M_{Y_{e}=0.5}$ by integrating the neutrino-driven mass loss
(eqs.~[\ref{eq:mdot_nu}] and [\ref{eq:mdot_nu2}]) across the
$Y_{e}^{a} \simeq 0.5$ region in Figure \ref{fig:mdot_radius}, using
$r_{d}(t)$ and $\dot{M}_{d}(t)$ from the disk evolution calculations
described in $\S$\ref{sec:results}.  

Pruet et al.~(2004) present calculations of $X_{\rm Ni}$ which are
parameterized in terms of the asymptotic entropy $S^{a}$, mass loss
rate $\dot{M}_{w}$, and asymptotic velocity $v^{a}$ of an outflow with
$Y_{e}^{a} \simeq 0.51$.  $M_{Y_{e} = 0.5}$ is dominated by
outflows from radii $\sim 3\times 10^{6}-10^{7}$ cm when
$\dot{M}_{d}\sim 0.1-1 M_{\sun}$ s$^{-1}$ (corresponding to $L_{52}
\sim$ few); equation (\ref{eq:sa}) thus gives $S^{a}\sim 10-30 k_{B}$
baryon$^{-1}$ for the ejecta with $Y_{e}^{a} \simeq 0.5$.  Purely
neutrino-driven winds achieve asymptotic velocities which are
typically below the escape speed of the central object (e.g., Thompson
et al.~2001); thus, the asymptotic kinetic energy is most likely
dominated by energy released during the formation of heavy elements.
Because $\sim 8$ MeV baryon$^{-1}$ is released in producing Fe-peak
elements, we estimate that $v^{a} \simeq 0.1-0.15$ c.  Applying these
wind parameters to Figure 3 of Pruet et al.~(2004), we estimate that $X_{\rm
Ni} \sim 0.2-0.5$, thereby justifying our scaling for $X_{\rm Ni}$ in
Figure \ref{fig:Nimass}.

Figure \ref{fig:Nimass} shows that for large initial disk masses
($M_{d,0} \gtrsim 0.1M_{\sun}$), the ejected Ni mass, $\sim 3 \times
10^{-4}-10^{-3}M_{\sun}$, can be appreciable.  Disks with moderate
initial radii $r_{d,0} \sim 10^{7}$ cm are optimal for producing
$^{56}$Ni because they are sufficiently large to contain the radius
$r_{\nu} \sim 10^{7}$ cm and yet are sufficiently compact to have a
large initial accretion rate, which maximizes the neutrino luminosity
and thus the neutrino-driven mass loss.  Conveniently, initial disk
parameters from many compact object merger simulations (see \S
\ref{sec:initial}) are in the range required to produce $\sim
10^{-4}-10^{-3}M_{\sun}$ of Ni.

\begin{figure}
\resizebox{\hsize}{!}{\includegraphics[ ]{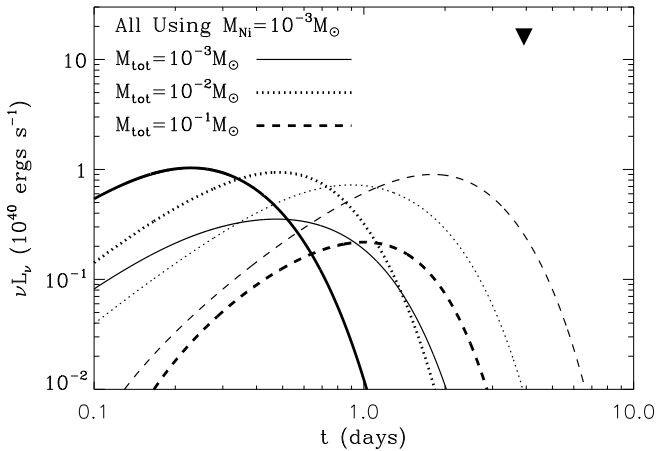}}
\caption{Luminosity of Ni decay-powered ``macronovae'' as a function
of time since merger for Ni mass $M_{\rm Ni} = 10^{-3}M_{\sun}$ and
ejecta velocity $v^{a} = 0.1$ c.  Light curves are shown for three
values of the total ejected mass $M_{\rm tot} = 10^{-3}$ ({\it solid
line}), $10^{-2}$ ({\it dotted line}), and $10^{-1}M_{\sun}$ ({\it dashed
line}).  The luminosities in V and J-Band (0.44 and 1.26 $\mu$m,
respectively) are shown with thick and thin lines, respectively.  The V-band upper limit on emission following GRB050509B from Hjorth et al.~(2005) is shown with a filled triangle.}
\label{fig:lightcurve}
\end{figure}

The decay of $M_{\rm Ni} \sim 10^{-4}-10^{-3}M_{\sun}$ can reheat the
(adiabatically cooled) ejecta sufficiently to produce detectable
transient emission.  In order to explore this possibility, we
calculate the light curves of ejecta heated by Ni decay
(``macronovae'') using the method of Kulkarni (2005).  This simplified
one-zone model accounts for the fraction of the gamma-rays produced by
the Ni decay which are absorbed by the expanding material (Colgate et
al.~1980) and assumes blackbody emission at the photosphere,
neglecting Comptonization.

Figure \ref{fig:lightcurve} shows the V and J-band luminosities as a
function of time since the merger for an outflow with Ni mass $M_{\rm
Ni} = 10^{-3}M_{\sun}$ which is expanding at $v^{a} = 0.1$ c.  The V-band light curve peaks earlier because the temperature at the photosphere decreases as the material expands.  Somewhat after the peak in the light curves, recombination will decrease the opacity well below that considered here; thus our calculations are not quantitatively reliable at these times.  The {\it total} mass $M_{\rm tot}$ ejected during the merger event, most of it
neutron rich, is likely to be significantly larger than $M_{\rm Ni}$;
this provides additional opacity for the Ni-rich material.  To explore
the effect of this additional material on the detectability of the Ni
decay, the light curves in Figure \ref{fig:lightcurve} are shown for
three values of $M_{\rm tot}$: $10^{-3} M_{\sun}$ ({\it solid line}),
$10^{-2}M_{\sun}$ ({\it dotted line}), and $10^{-1}M_{\sun}$ ({\it dashed
line}).  As Figure \ref{fig:lightcurve} shows, larger $M_{\rm tot}$:
(1) delays the time to peak emission ($t_{\rm peak}$ is roughly
$\propto M_{\rm tot}^{1/2}$); (2) increases the total fluence of the
event by trapping a higher fraction of the gamma-ray emission; and (3)
increases the peak wavelength of the emission, pushing it into the
near-IR for large $M_{\rm tot}$.  We conclude that long wavelength
($\lambda \gtrsim \mu$m) observations at $t \sim 1$ day are the most
promising for the detection of a Ni decay-powered macronova.

Hjorth et al.~(2005) place an upper limit of $M_{V} > 27.5$ at $t
=3.9$ days on any emission associated with the short GRB 050509B
(redshift $z \simeq 0.22$); we mark this constraint in Figure
\ref{fig:lightcurve} with an arrow.  For $M_{\rm tot} = 0.1M_{\sun}$
this constrains the ejected Ni mass to be $M_{\rm Ni} \lesssim
10^{-2}M_{\sun}$ (see also Kulkarni 2005).  As Figure \ref{fig:Nimass}
illustrates, compact object mergers are very unlikely to produce this
much Ni, so the absence of a detection thus far is unsurprising.

\section{Conclusion and Discussion}
\label{sec:theend}

We have calculated the time-dependent evolution of accretion disks
formed from compact mergers, and the properties of their outflows.
Since most of the disk mass resides at large
radii, we approximate the disk as a ring at a given radius and
calculate the dynamics and composition of the ring as a function of
time. This ring model is calibrated to correctly reproduces
the Green's function solution for a viscously spreading ring with
viscosity $\nu \propto r^{1/2}$ (appropriate for a thick disk; see
Appendix \ref{appendix:ring}). With this simplified model, we have
studied the full parameter space of remnant accretion disks (different
initial masses, compositions, etc.) and can follow the viscous
evolution for arbitrarily long timescales.

The energetics of the ring at a given time can be described by one of
three models: (1) optically thick to neutrinos and advective, (2)
optically thin to neutrinos and geometrically thin, and (3) optically
thin to neutrinos and advective. A massive, compact disk (with a short
initial viscous time $t_{\rm visc,0}$; eq. [\ref{eq:tvisc0}]) will
exhibit all three of these accretion phases, evolving from (1) to (3)
as a function of time (Figs. \ref{fig:mass1}-\ref{fig:energy2}).
Less massive disks, on the other hand, only pass through phases
(2) and (3), or even just (3). Note that these phases refer to the
energetics of the disk near the outer radius.  At a given time, the disk may also undergo similar transitions as a
function of radius; e.g., a disk that is advective at large radii will
be neutrino cooled and geometrically thin inside the ignition radius
$r_{\rm ign}$ (eq. [\ref{eq:rign}]).

Neutrino-driven winds during the early-time optically thick and
neutrino-cooled (thin disk) phases unbind so much mass that field
lines connected to the disk cannot produce sufficiently relativistic
material to power short-duration GRBs (\S \ref{thindiskwinds} and
Fig. \ref{fig:mdot}).  An alternative source for the relativistic
material needed to produce short GRBs are nearly baryon-free magnetic
field lines that thread the BH's event horizon (e.g., McKinney 2005).
In addition, when the inner disk becomes advective ($\dot M_{d} \lesssim
0.07 \, \alpha_{0.1}^{5/3} \, M_\odot \, s^{-1}$ for a = 0),
conditions appear particularly suitable for the formation of
relativistic jets (by analogy to X-ray binaries, which produce jets
when making a similar transition; e.g., Remillard $\&$ McClintock
2006; see Lazatti et al.~2008 for a similar argument in the context of long-duration GRBs).

Once the disk has transitioned to a late-time advective phase (phase 3
above), the properties of the disk become well-described by
self-similar solutions.  Ignoring for the moment outflows from the
disk, these solutions are $r_d \propto t^{2/3}$, $M_d \propto
t^{-1/3}$, and $\dot M_{d} \propto t^{-4/3}$.  Power-law variations in the
disk properties are a generic feature of a viscously evolving disk
that conserves total angular momentum.  These scalings are not,
however, likely to be applicable in practice because outflows during
the advective phase unbind most of the remaining material (\S
\ref{sec:gross} \& \ref{thickdiskwinds}).  Energy produced by fusion
to He and heavier elements also contributes to driving an outflow
(Figs. \ref{fig:energy1} \& \ref{fig:energy2}).  Such outflows remove a significant fraction of the angular momentum of the disk.
This leads to a much more rapid decrease in the disk mass and accretion rate at
late times (Appendix B3 and Fig. \ref{fig:wind}).  Significant
accretion onto the central black hole will thus only last for a few
viscous times after the onset of the advective phase.

At the outer edge of the disk, the transition from a neutrino-cooled
thin disk to the late-time advective phase occurs at a time $t_{\rm
thick} \sim 0.1 \, \alpha_{0.1}^{-23/17} \, (J_{\rm 49}/2)^{9/17}$ s
(eq. [\ref{eq:tthick}]).  The rapid decrease in $\dot M_{d}$ after the
onset of the advective phase implies that the inner disk becomes
advective at a similar time (\S \ref{thickdiskwinds} and
Fig. \ref{fig:wind}).  Quantitatively, we find that for powerful winds
with $p = 1$ (see eq. [\ref{outflow}]), the inner disk becomes
advective at $t \sim 0.2$, $5$, and $100$ sec, for $\alpha = 0.1,
0.01,$ and $0.001$, respectively (for our fiducial model with an
initial mass of $0.1 M_\odot$ and an initial radius of $\simeq 3
\times 10^7$ cm).  Thus, for $\alpha \sim 10^{-3}$, the timescale for
the inner disk to become advective is comparable to the onset of
observed flaring at $\sim 30$ sec in some short GRBs (e.g., Berger et
al. 2005).  Given the slow decline in disk mass with time before
$t_{\rm thick}$, there is ample accretion energy available in the disk
at this point to power the observed flaring.  However, there is
observational evidence for $\alpha \sim 0.1$ in a number of
astrophysical disks (King et al. 2007); we thus doubt that $t_{\rm
thick}$ is large enough to coincide with the onset of observed
flaring.  Instead $t_{\rm thick}$ is likely to be $\sim 0.1-1$ sec,
comparable to the duration of the short GRB itself. In this case, the
rapid decrease in the disk mass and accretion rate in the advective
phase imply that the remnant accretion disk alone does not contain
sufficient mass at $\sim 30$ sec to power the observed late-time
activity from short GRBs, nor is there any physical reason to expect a
sudden change in the disk or jet properties at this time.  

A more likely source of late-time flaring in compact object merger models is a continued inflow of mass at late
times, such as is produced by the infalling tidal tail found in Lee \&
Ramirez-Ruiz's (2007) NS-NS merger simulations (see also Rosswog 2007).  Similarly, the BH-NS merger simulations of Faber et al.~(2006a,b) show that $\sim 0.03 M_{\sun}$ of material is ejected into highly eccentric orbits during the merger, which returns to the BH on a timescale $\gtrsim 1$ s.  However, final conclusions regarding the quantity and ubiquity of late-time fall-back from NS-NS and BH-NS mergers must await full-GR simulations which include BH spin and realistic EOSs.

The second major focus of this paper has been on the composition of
the accretion disk and its outflows as a function of time.  For
initial disk properties expected in compact object mergers (\S
\ref{sec:initial}), the disk typically comes into $\beta$-equilibrium
given the high temperatures and densities at small radii.  As material
spreads to larger radii, however, the composition of the disk freezes
out before it becomes advective at late times; at freeze-out the disk
is modestly neutron rich, with an electron fraction $Y_e \approx 0.3$ (\S \ref{sec:composition} and Fig. \ref{fig:Ye}).  This
neutron rich material -- $\sim 10^{-2} M_\odot$ for typical initial
disk parameters -- is blown away once the disk enters the advective
phase at $\sim t_{\rm thick}$.  These outflows are particularly
interesting given the low solar system abundance of material produced
in nuclear statistical equilibrium at $Y_e \sim 0.3$ (Hartmann et
al. 1985).  In a separate paper, we will study this nucleosynthesis
and its implications in more detail.

Although outflows from compact object merger accretion disks are
neutron rich in most circumstances, neutrino-driven winds from radii
$\simeq 10^6-10^7$ cm at accretion rates $\dot M_{d} \sim 0.03-1 \,
M_\odot \, {\rm s^{-1}}$ have electron fractions $Y_e \simeq 0.5$,
precisely that required to synthesize significant amounts of $^{56}$Ni
(Fig. \ref{fig:mdot_radius}). We have calculated the total Ni mass
ejected by compact object merger disks as a function of their initial
mass and radius (\S \ref{sec:ni} and Fig. \ref{fig:Nimass}). Disks
with initial masses $\gtrsim 0.1 M_\odot$ can produce up to $\sim
10^{-3} M_\odot$ of $^{56}$Ni.  The radioactive decay of this Ni as
the outflow expands to large radii will produce an optical and
infrared transient peaking $\sim 0.5-2$ days after the merger, with a
peak flux of $\nu L_\nu \simeq 10^{40}$ ergs s$^{-1}$
(Fig. \ref{fig:lightcurve}).  Because the Ni mass is likely to be a
small fraction of the total mass of the ejecta (most of which is
neutron rich), this transient is best detected at $\sim 1 \, \mu$m.
As Figure \ref{fig:lightcurve} shows, current observational limits on
SN-like transients coincident with short GRBs are about a factor of
$\sim 10$ above our predictions.  However, somewhat deeper limits from
a moderately closer burst could start to put interesting constraints
on short GRB progenitors.  It is also possible that the decay of some
neutron-rich isotopes could heat the outflow and contribute to the
late-time thermal emission (although most such isotopes have very
short half-lives).  This possibility should be investigated in future
calculations using a nuclear reaction network.

Although we have focused on short GRBs throughout this paper, many of
our results can be applied more broadly.  For example, long duration
GRBs show late-time activity and flaring similar to that seen in short
GRBs (e.g., Falcone et al.~2007).  For the reasons described above, this activity is
probably produced by a continued inflow of mass at late times
(fallback from the stellar progenitor's envelope) rather than solely
by the viscous evolution of the small-scale disk.  As a final
application of our results, we note that the accretion-induced
collapse of a white dwarf to a neutron star (AIC) is expected to
produce a compact disk of $\sim 0.1-0.5 M_{\sun}$ outside the newly
formed neutron star's surface (Dessart et al. 2006).  The calculations
presented here describe the evolution of this remnant disk, with the
one caveat that the composition of the disk in the AIC context may be
strongly affected by neutrino irradiation from the newly-formed
neutron star.

\section*{Acknowledgments}

We thank Josh Bloom, Davide Lazzati, and Daniel Perley for useful conversations.  A. L. P. is supported by the Theoretical Astrophysics Center at UC
Berkeley.  B. D. M.  and E. Q. are supported in part by the David and
Lucile Packard Foundation, NASA Grant NNG06GI68G, and a NASA GSRP
Fellowship to B.D.M.

\begin{appendix}

\section{Calibration of the Ring Model}
\label{appendix:ring}

The surface density $\Sigma$ of an axisymmetric disk in a Keplerian
potential with constant total angular momentum evolves according to a
diffusion equation (e.g., Frank et al.~2002): \be \frac{\partial
\Sigma}{\partial t} = \frac{3}{r}\frac{\partial}{\partial
r}\left[r^{1/2}\frac{\partial}{\partial r}\left(\nu\Sigma
r^{1/2}\right)\right],
\label{eq:sigma_evo}
\ee where $\nu$ is the kinematic viscosity.  Assuming that $\nu$
depends only on radius as a power law, viz: $\nu =
\nu_{0}(r/R_{0})^{n}$, equation (\ref{eq:sigma_evo}) is linear and,
for an initial surface density distribution $\Sigma(r,t=0) =
(M_{0}/2\pi R_{0})\delta(r-R_{0})$ which is narrowly peaked about the
radius $R_{0}$, the solution (for $n < 2$) is given by \bea \Sigma(r,t) = \nonumber \eea \be
\frac{M_{0}(1-n/2)}{\pi
R_{0}^{2}x^{(n+1/4)}\tau}\exp\left[\frac{-(1+x^{2-n})}{\tau}\right]I_{1/|4-2n|}\left[\frac{2x^{1-n/2}}{\tau}\right], \nonumber \\
\label{eq:sigma}
\ee where $M_{0}$ is the initial disk mass, $x \equiv r/R_{0}$, $\tau
\equiv t[12\nu_{0}(1-n/2)^{2}/R_{0}^{2}]$, and $I_{m}$ is a modified
Bessel function of order $m$.  For small argument $y \ll 1$,
$I_{m}(y)$ takes the asymptotic form $I_{m} \simeq
(y/2)^{m}/\Gamma(m+1)$, where $\Gamma$ is the Gamma function; thus,
for late times or small radii such that $\tau \gg 2x^{1-n/2}$,
equation (\ref{eq:sigma}) reduces to \bea \Sigma(r,t)|_{\tau \gg
2x^{1-n/2}} = \nonumber \eea \bea \frac{M_{0}}{\pi
R_{0}^{2}}\frac{(1-n/2)}{\Gamma[\frac{5-2n}{4-2n}]}\frac{1}{\tau^{\left(\frac{5-2n}{4-2n}\right)}x^{n}}\exp\left[\frac{-(1+x^{2-n})}{\tau}\right]
\label{eq:sigma_asym}
\eea Most of the mass in the disk is located near the radius where the
local mass $M_{d} \propto \Sigma r^{2}$ peaks; using equation
(\ref{eq:sigma_asym}), at late times this radius is found to be
$r_{\rm peak} = R_{0}\tau^{1/(2-n)}$.  Hence, equation
(\ref{eq:sigma_asym}) becomes valid near $r_{\rm peak}$ for $\tau \gg
1$.

The constant $A$, which relates the total disk mass at late times from
the exact solution of equation (\ref{eq:sigma_evo}) to the mass
defined by $\pi\Sigma(r_{\rm peak})r_{\rm peak}^{2}$, can
be calculated from equation (\ref{eq:sigma_asym}) to be \be A(\tau \gg
1) \equiv \left.\frac{\int_{0}^{\infty}2\pi \Sigma r
dr}{\pi\Sigma(r_{\rm peak})r_{\rm peak}^{2}}\right|_{\tau \gg 1} =
\frac{2e}{2-n}
\label{eq:a_const}
\ee Similarly, the constant $B$, which relates the total disk angular
momentum at late times from the exact solution to that estimated by
$\pi\Sigma r_{\rm peak}^{2}(GMr_{\rm peak})^{1/2}$, is
given by \be B(\tau \gg 1) \equiv \left.\frac{\int_{0}^{\infty}2\pi
\Sigma r^{3/2} dr}{\pi\Sigma(r_{\rm peak})r_{\rm
peak}^{5/2}}\right|_{\tau \gg 1} =
\frac{2e}{2-n}\Gamma\left[\frac{5-2n}{4-2n}\right]
\label{eq:b_const}
\ee 

From mass continuity, the radial velocity is given by
\be
v_{r} = \frac{-3}{\Sigma r^{1/2}}\frac{\partial}{\partial r}\left[\nu\Sigma r^{1/2}\right] = \frac{-3\nu_{0}}{R_{0}}\frac{1}{\Sigma x^{1/2}}\frac{\partial}{\partial x}\left[\Sigma x^{n+1/2}\right],
\label{eq:velocity}
\ee
which, using equation (\ref{eq:sigma_asym}), gives the accretion rate at small radii
\bea \dot{M}_{\rm in} &=& -2\pi\Sigma r v_{r}|_{\tau \gg 2x^{1-n/2}} \nonumber \\ &=& \frac{M_{0}}{R_{0}^{2}/\nu_{0}}\frac{3(1-n/2)}{\Gamma[(5-2n)/(4-2n)]}\exp[-1/\tau]\tau^{-\left(\frac{5-2n}{4-2n}\right)} \nonumber \\
\label{eq:mdot_analytic}
\eea
Equation (\ref{eq:mdot_analytic}) is easily checked by noting that $\int_{0}^{\infty}\dot{M}_{\rm in}dt = M_{0}$, which shows that the entire initial disk eventually accretes onto the central object.  In $\S\ref{sec:dynamical}$ we introduced the following prescription for evolving the disk mass:
\be
\dot{M}_{d} = \frac{fM_{d}}{t_{\rm visc}},
\label{eq:mdot_f}
\ee where, in terms of the viscosity prescription adopted above,
$t_{\rm visc} = r_{d}^{2}/\nu = t_{\rm visc,0}(r_{d}/R_{0})^{2-n}$ and
$t_{\rm visc,0} \equiv R_{0}^{2}/\nu_{0}$ is the initial viscous time.
Assuming that the total disk angular momentum remains constant, $J
\propto M_{d}r_{d}^{1/2} = M_{0}R_{0}^{1/2}$, the solution to equation
(\ref{eq:mdot_f}) is given by \be M_{d}(t) = M_{0}[1 +
(4-2n)f(t/t_{\rm visc,0})]^{-1/(4-2n)}
\label{eq:mdisk}
\ee
In our evolutionary calculations we set $f$ so that the accretion rate from the exact solution to equation (\ref{eq:sigma_evo}) ($\dot{M}_{\rm in}$; eq.~[\ref{eq:mdot_analytic}]) matches the solution to equation (\ref{eq:mdot_f}) at late times (i.e., in the self-similar limit).  This requires
\be
f = 3(1-n/2)\Gamma[(5-2n)/(4-2n)]^{4-2n}
\label{eq:fspecial}
\ee

For an advection-dominated disk, $\nu = \alpha c_{s}H \propto \Omega R^{2} \propto r^{1/2}$; thus, $n=1/2$, $f \simeq 1.602$, $A \simeq 3.62$,
and $B \simeq 3.23$.  For a neutrino-cooled, optically-thin disk which
is dominated by gas pressure, $T \propto r^{-3/10}$ and $\nu \propto
r^{6/5}$; thus, $n=6/5$, $f \simeq 1.01$, $A \simeq 6.80$, and $B
\simeq 6.09$.

\begin{figure}
\resizebox{\hsize}{!}{\includegraphics[]{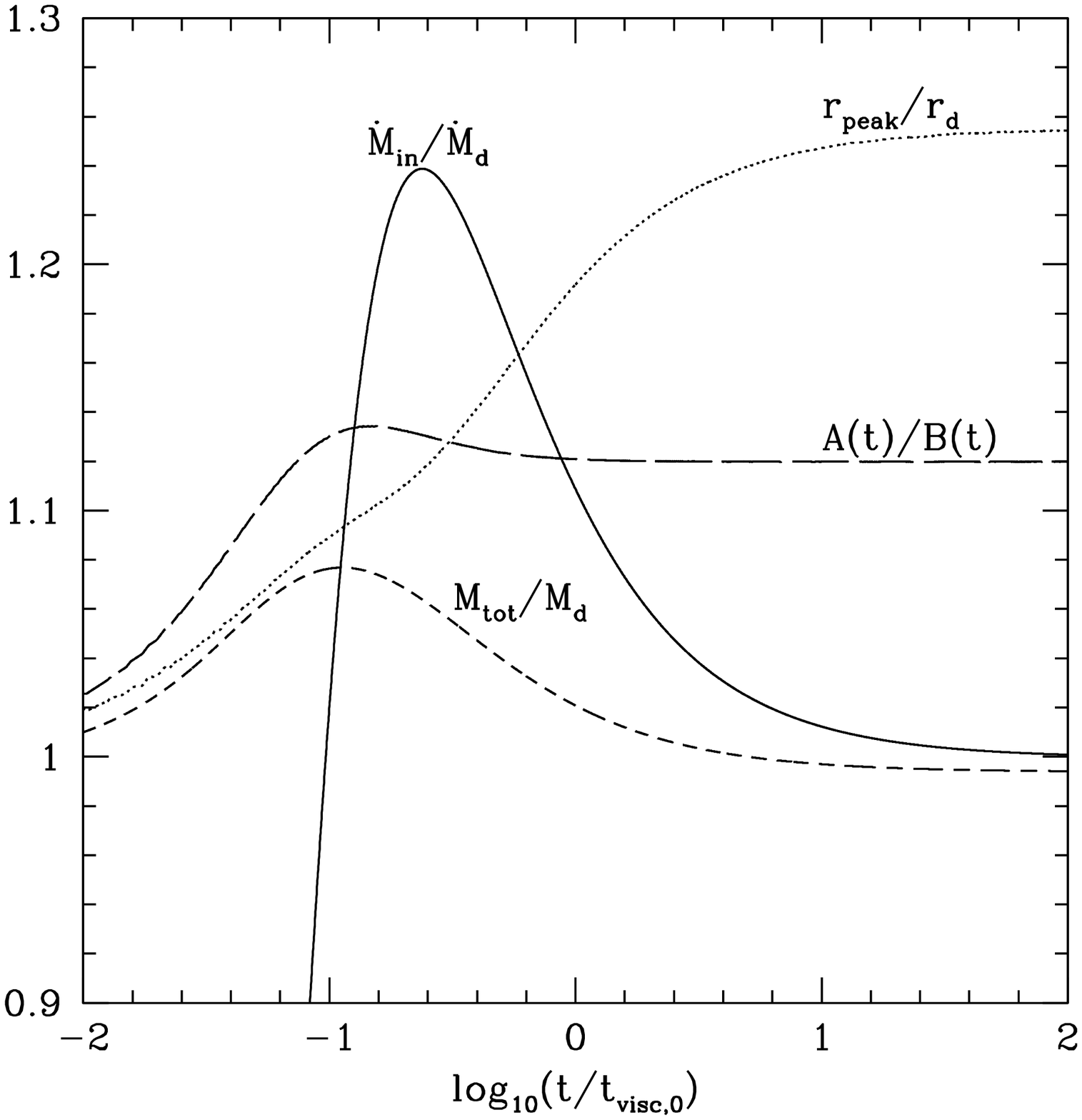}}
\caption{Comparison of the accretion rate ($solid$), disk mass
($short$ $dashed$), and disk radius (where the local disk mass peaks;
$dotted$) as calculated from our simplified ring model to that derived
from the exact solution of the diffusion equation for a
$\delta-$function initial mass distribution (eq.~[\ref{eq:sigma}]); we
assume $\nu \propto r^{1/2}$, as applies for a thick disk.  The
parameter $f \simeq 1.6$ (eq.~[\ref{eq:fspecial}]) adopted in our
model is chosen to ensure that the accretion rates match at late times
(i.e., $\dot{M}_{\rm in}/\dot{M}_{d} \rightarrow 1$).  Also shown is
the ratio $A(t)/B(t)$ (eqs.~[\ref{eq:a_const}] and
[\ref{eq:b_const}]), a measure of the relative distribution of mass
and angular momentum, which asymptotes to $\Gamma[(5-2n)/(4-2n)]
\simeq 1.12$ at late times.}
\label{fig:comparison}
\end{figure}

In Figure \ref{fig:comparison} we show $\dot{M}_{\rm in}/\dot{M}_{d}$
as a function of $t/t_{\rm visc,0}$ for $n=1/2$ in order to compare
the disk evolution derived from the exact solution of equation
(\ref{eq:sigma_evo}) to that calculated from our simplified model.
Figure \ref{fig:comparison} also shows the ratio of the total disk
mass $M_{\rm tot} \equiv \int_{0}^{\infty}2\pi \Sigma r dr$ calculated
from equation (\ref{eq:sigma}) to the disk mass $M_{d}$
(eq.~[\ref{eq:mdisk}]) of the simplified model, as well as the ratio
of $r_{\rm peak}$ (the radius where $\Sigma r^{2}$ peaks, using
eq.~[\ref{eq:sigma}] for $\Sigma$) to the radius determined by angular
momentum conservation: $r_{d} = R_{0}(M_{d}/M_{0})^{2}$.  Figure
\ref{fig:comparison} shows that, although the accretion rate in the
two models differ at very early times (the initially
narrowly-concentrated ring takes a short period of time to spread to
small radii), they approach one another to $\lesssim 20\%$ by $t
\gtrsim 0.1t_{\rm visc,0}$.  Likewise, the disk mass and radii from
the exact solution and simplified model are quite similar at all
times.

The numerical values for $A$ and $B$ given in equations
(\ref{eq:a_const}) and (\ref{eq:b_const}) and employed in our
calculations apply only to the mass and angular momentum distribution
in the disk at late times ($\tau \gg 1$).  Initially, the disk is
entirely concentrated at a single radius and $A(t = 0) = B(t = 0) =
1$; thus, $A(t)$ and $B(t)$ evolve significantly from early times
until the disk enters the self-similar limit and so one might worry
that the early-time description of the disk's evolution depends
sensitively on the initial mass distribution.  Our model only assumes,
however, that the $ratio$ $A(t)/B(t)$ remains constant, which is a
good approximation.  To illustrate this, Figure \ref{fig:comparison}
shows $A(t)/B(t)$ calculated from the exact solution
(eq.~[\ref{eq:sigma}]) for $n = 1/2$.  Note that $A(t)/B(t)$ increases
from unity to its asymptotic value $A/B = \Gamma[(5-2n)/(4-2n)]$,
which is $\simeq 1.12$ for $n = 1/2$.

\section{Analytic Self-Similar Solutions}
\label{sec:analytic}

The late-time evolution of our disk calculations asymptote to power
laws that are well approximated by analytic self-similar solutions.
We derive these here to aid in interpreting our numerical
results. Presentation is divided between neutrino-cooled, thin-disk
solutions and late-time advective solutions. One could just as well
derive analogous results for disks that are optically thick to
neutrinos. We forgo this here since the initial viscous time is always
sufficiently long that these solutions are never applicable to our
numerical results.  We conclude by presenting self-similar solutions
for advective disks with substantial mass loss, since these differ
significantly from the solutions without mass loss.

\subsection{Neutrino-cooled, Thin-disk Solutions}

\begin{figure}
\resizebox{\hsize}{!}{\includegraphics[ ]{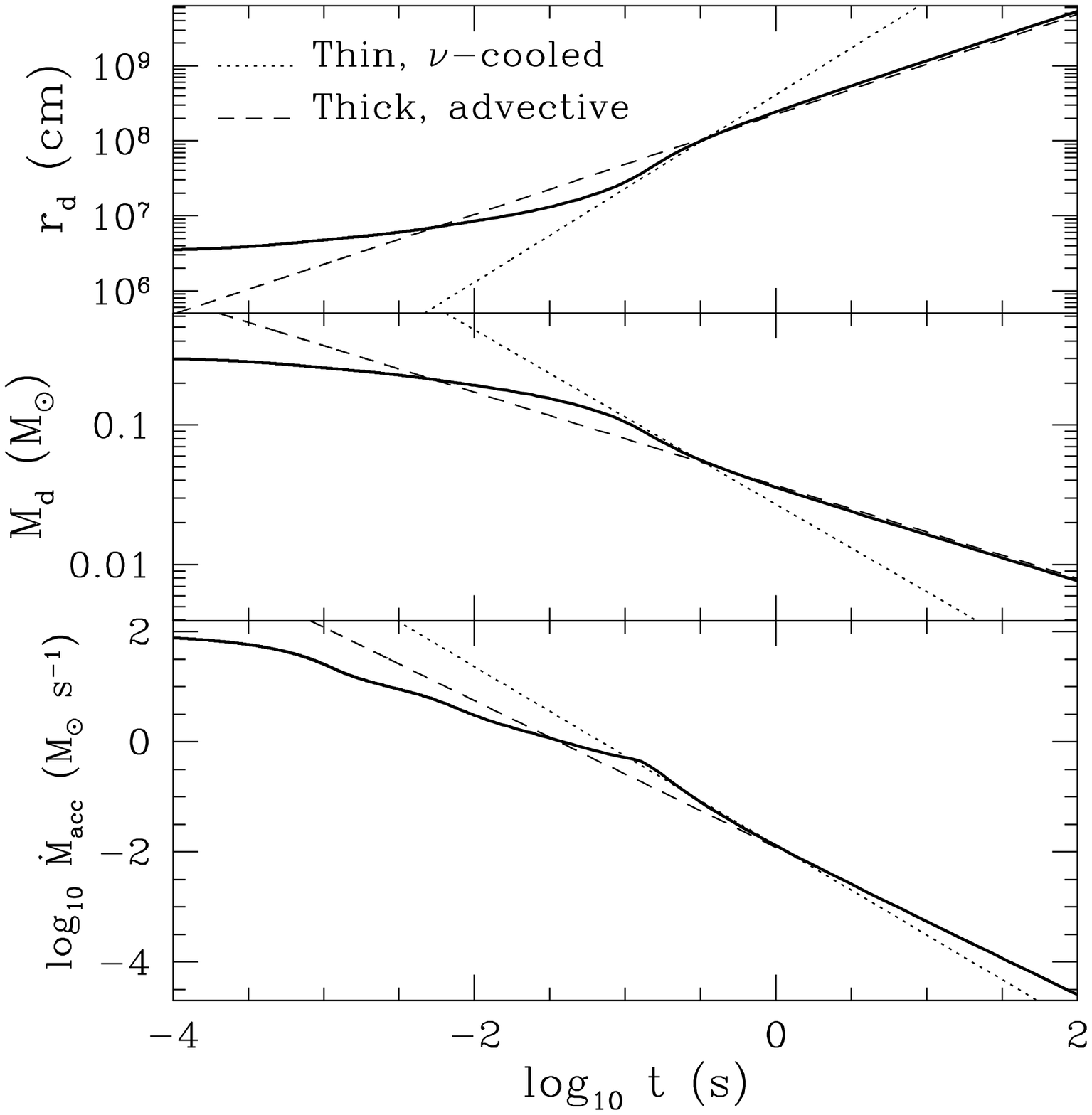}}
\caption{Comparison of the numerical disk solutions ({\it solid lines}) with
the analytic solutions for the thin, neutrino-cooled ({\it dotted lines})
and thick, advective limits ({\it dashed lines}). The numerical solution is the
$0.3M_\odot$ disk from Fig. \ref{fig:mass1}.}
\label{fig:analytic}
\end{figure}

In the neutrino-cooled, thin-disk limit, the cooling is dominated by
Urca, and the pressure is given by ideal gas. Combining local energy
balance and continuity, $\dot{M}_{d}=fA\pi\nu\Sigma$, allows us to
solve for the temperature and column density as functions of radius.
We substitute these into the angular momentum equation,
$B(GMr_d)^{1/2}\pi r_d^2\Sigma=J_d$, to solve for $M_d$ as a function
of $\dot{M}_{d}$ and $J_d$. We then assume the solutions have a
self-similar form of $M_d\propto t^{-\beta}$, so that
$\dot{M}_{d}=-dM_d/dt=\beta M_d/t$. In this way we solve for
$\beta=5/8$, $\dot{M}_{d}(t)$, and subsequently any other variable of
interest. The results are \bea M_d= \nonumber \eea \bea 1.3\times10^{-2}
f_{1.6}^{-5/8}\left(\frac{A_{3.6}}{B_{3.2}}\right)\alpha_{0.1}^{-3/4}M_3^{-1/4}
\left(\frac{J_{49}}{2}\right) t^{-5/8} M_\odot,
	\label{eq:md_analytic1}
\eea
\bea
	\dot{M}_{d}= \nonumber
\eea \bea
2.7\times10^{-2}f_{1.6}^{-5/8}\left(\frac{A_{3.6}}{B_{3.2}}\right)\alpha_{0.1}^{-3/4}M_3^{-1/4}
		\left(\frac{J_{49}}{2}\right) t^{-13/8}M_\odot{\rm s^{-1}}
	\label{eq:mdot_analytic1}
\eea
and
\be
	r_d=4.1\times10^8f_{1.6}^{5/4}\alpha_{0.1}^{3/2}M_3^{-1/2}
	t^{5/4}{\rm cm}.
	\label{eq:rd_analytic1}
\ee where $f_{1.6}=f/1.6$, $A_{3.6}=A/3.6$, $B_{3.2}=B/3.2$, and $t$
is measured in seconds, and the prefactors have been scaled to match
our numerical results.  The first thing to notice is that both $M_d$
and $\dot{M}_{d}$ are rather insensitive to the choice of $f$ as long
as it is near unity, and $A$ and $B$ only appear as a ratio, which is
also nearly unity. This provides confidence in using this
parameterization, and these specific values for the corresponding
parameters, when the disk is not well-described by $n=1/2$. This
analysis also demonstrates the relative dependence on $\alpha$.  In
Figure \ref{fig:analytic} we compare these scaling ({\it dotted
lines}) with the numerical calculations. This shows that these
solutions are only applicable for a short time.  At times when
$t<t_{\rm visc}$ the evolution is much flatter and is dominated by
initial conditions. At later times the disk becomes advective and the
solutions of the next section apply.

\subsection{Late-time Advective Solutions}

\label{sec:riaf}

In this limit, self-similar solutions can be found in an analogous way. The viscous
energy release is carried by advection with the internal energy dominated
by relativistic particles, so that
\be
	\frac{9}{8fA\pi}\Omega^2\dot{M}=V_r\frac{H}{r}\frac{11}{6}aT^4.
	\label{eq:appendix_energy}
\ee Combining this with mass continuity, gives the column depth as a
function of radius, $\Sigma(r)=(16/9A\pi\alpha)(\dot{M}/r^2\Omega)$.
We then use this relation with $B(GMr_d)^{1/2}M_d=J_d$ and
$\dot{M}_{d}=\beta M_d/t$, to find $\beta=1/3$ and the self-similar
solutions \be M_d =
3.7\times10^{-2}\left(\frac{A_{3.6}}{B_{3.2}}\right)\alpha_{0.1}^{-1/3}M_3^{-2/3}
\left(\frac{J_{49}}{2}\right) t^{-1/3}M_\odot,
	\label{eq:md_analytic2}
\ee
\be
	\dot{M}_{d}=1.2\times10^{-2}\left(\frac{A_{3.6}}{B_{3.2}}\right)\alpha_{0.1}^{-1/3}M_3^{-2/3}
		\left(\frac{J_{49}}{2}\right) t^{-4/3}M_\odot{\rm s^{-1}},
	\label{eq:mdot_analytic2}
\ee
and
\be
	r_d=2.3\times10^8\alpha_{0.1}^{2/3}M_3^{1/3}
	t^{2/3}{\rm cm}.
	\label{eq:rd_analytic2}
\ee
These advective results are even more insensitive to $A$, $B$, and $f$ than the
thin-disk results. Equation (\ref{eq:md_analytic2})-(\ref{eq:rd_analytic2})
are plotted in Figure \ref{fig:analytic} as dashed lines. The numerical calculations follow
these solutions very closely for times later than $t_{\rm thick}$
(given by eq. [\ref{eq:tthick}]).

   Equations (\ref{eq:md_analytic2})-(\ref{eq:rd_analytic2}) can also
be derived ignoring equation (\ref{eq:appendix_energy}), but assuming
that the scaleheight is fixed at $H/r\simeq0.6$. This introduces the
additional dependencies $M_d\propto(H/r)^{-2/3}$,
$\dot{M}_{d}\propto(H/r)^{-2/3}$, and $r_d\propto(H/r)^{4/3}$, but
gives nearly identical prefactors.

\subsection{Advective Solutions with Mass Loss}

\label{sec:riaf}

In \S \ref{thickdiskwinds} we described how advective disks are likely
to lose a substantial fraction of their mass to viscously driven
outflows.  Because the outflow removes angular momentum as well -- at
least the specific angular momentum of the mass that is lost -- the
disk need not expand as rapidly to large radii.  In addition, the disk
mass and accretion rate decrease much more rapidly at late times than
in the self-similar solutions described in the previous subsection.
To quantify this effect, we follow Blandford \& Begelman (1999) and
assume that only a fraction $\sim (r_*/r_d)^p$ of the available
material is accreted onto the central BH.  The remainder is lost to an
outflow.  Thus the outflow rate at any time is given by \be \dot
M_{\rm out} = \left( 1 - \left[r_* \over r_d\right]^p\right) \, {f M_d \over
t_{\rm visc}} \label{outflow} \ee We further assume that the angular
momentum loss rate from the disk is given by \be \dot J = - C \dot
M_{\rm out} \left( G M r_d\right)^{1/2}
\label{ang}.  \ee where $C$ is a constant that depends on the torque
exerted by the outflowing mass on the remaining disk.  If the outflow
produces no net torque, an assumption that appears at least
qualitatively consistent with the relatively small-scale magnetic
fields seen in global MHD disk simulations (e.g., Stone \& Pringle
2001), then the angular momentum loss is only that due to the specific
angular momentum of the outflow, and (Kumar, Narayan, \& Johnson 2008)
\be C = {2p \over 2p + 1} \label{C}.  \ee We solve equations
(\ref{eq:mass}), (\ref{eq:ang_mom}), (\ref{outflow}), and (\ref{ang}),
assuming $A/B = 1$ and $\nu \propto r^{1/2}$ (as appropriate for a
thick disk).  The solution depends on the relative magnitude of $1-C$
and $C(r_*/r_d)^p$.  For $C(r_*/r_d)^p \ll 1-C$, which is true at
nearly all times if equation (\ref{C}) is applicable, then \be r_d
\simeq r_{d,0} \left[1 + 3 f (1-C)\left({t \over t_{\rm
visc,0}}\right)\right]^{2/3}, \label{rdCne1} \ee \be M_d \simeq M_{d,0}
\left[1 + 3 f (1-C) \left({t \over t_{\rm visc,0}}\right)\right]^{-1/ 
[3(1-C)]},
\label{MdCne1}\ee and \bea \dot M_{\rm in} \simeq {f {M_{d,0} \over t_{\rm visc,0}}}\left(r_* \over r_{d,0}\right)^p\times \nonumber \eea
\bea 
\left[1 + 3 f (1-C) \left({t \over t_{\rm visc,0}}\right)\right]^{-[1
+ 3(1+2p/3)(1-C)]/[3(1-C)]}
\label{mdotinCne1} \eea Note that if $p = C = 0$ (i.e., no mass or
angular momentum loss), then these self-similar solutions reduce to
those of the previous subsection.  However, for the case $p = 1$
consistent with a number of global advective disk simulations (e.g.,
Hawley \& Balbus 2002), and in the absence of a net torque on the
disk, $C = 2/3$ and these solutions correspond to $r_d \propto
t^{2/3}$, $M_d \propto t^{-1}$, and $\dot M_{\rm in} \propto t^{-8/3}$
(see also Fig. \ref{fig:wind}).  This shows that the disk mass and
accretion rate decrease subsantially more rapidly in time than in the
absence of an outflow, while the disk expands outward at roughly the
same rate.  If there is a net torque on the disk such that $C \simeq
1$, then equations (\ref{rdCne1})-(\ref{mdotinCne1}) are not
applicable.  Instead, for $C(r_*/r_d)^p \gg 1-C$, the solution is
given by (for $p \ne 0$ and $t \gg t_{\rm visc,0}$) \be r_d(t) \simeq
\left[(3 + 2p) f r_*^p r_{d,0}^{1.5}\right]^{1/(1.5 + p)} \, \left( t
\over t_{\rm visc,0}\right)^{1/(1.5 + p)} \label{rd} \ee and \be M_d(t)
\simeq M_{d,0} \exp[-D (t/t_{\rm visc,0})^{p/(1.5 + p)}]
\label{Md} \ee
where \be D = \left( {1.5 + p \over p \, (3 + 2 p)^{1.5/(1.5 + p)}} \right)
\left( {f r_{d,0}^{1.5} \over [f r^p_{*} r_{d,0}^{1.5}]^{1.5/(1.5 +
p)}}\right).  \ee For $p = 1$ and for $r_{d,0} \sim r_*$, these solutions
become $r_d(t) \sim r_{d,0} (t/t_{\rm visc,0})^{2/5}$ and $M_d(t) \sim
M_{d,0} \exp[-1.15 (t/t_{\rm visc,0})^{2/5}]$.  The radius of the disk
thus increases significantly more slowly, and the mass of the disk
decreases much more rapidly, than in the self-similar solutions
without mass-loss.

The numerical solutions including mass-loss during the advective phase
shown in \S 4.1 (Fig. \ref{fig:wind}) assume that equation (\ref{C})
is applicable and are indeed well-described by the self-similar
solutions given in equations (\ref{rdCne1})-(\ref{mdotinCne1}) at late
times.

\end{appendix}

\label{lastpage}

\end{document}